\documentclass[10pt,sigconf,letterpaper]{acmart}

\setcopyright{rightsretained}
  \acmConference[draft (under submission)]{draft (under submission)}{June 2020}{Marina del Rey, CA, USA}
\acmYear{2020}
\copyrightyear{2020}
\acmISBN{}
\acmDOI{}

\usepackage{tikz}
\usepackage{amsmath}

\usepackage{multirow}
\usepackage{url}
\usepackage{multicol}
\usepackage{subcaption}
\usepackage{graphicx} 
\usepackage{hyperref}
\usepackage{amsmath}
\usepackage{color, colortbl}
\usepackage{xspace}
\definecolor{Gray}{gray}{0.9}
\usepackage{tikz} 
\usetikzlibrary{arrows,automata}
\usepackage[title]{appendix}

\hypersetup{colorlinks,linkcolor=[rgb]{.75,0,0},urlcolor=[rgb]{0.75,0,0.75},citecolor=blue,breaklinks} %

\newcommand\V[1]{{\mbox{\textit{#1}}}}
\newcommand\Vs[1]{{\mbox{\small \textit{#1}}}}
\newcommand\R[1]{{\mbox{{#1}}}}

\makeatletter
  \newcommand\EatSpacesHack{\@bsphack\@esphack}
\makeatother
  \renewcommand\comment[1]{\EatSpacesHack}
  \newcommand\reviewfix[1]{\EatSpacesHack}
  \newcommand\PostSubmission[1]{\EatSpacesHack}
\makeatother

\def\Snospace~{\S{}} %
\date{}

  \newcommand{\broot}{\texttt{B-root}\xspace}
  \newcommand{\brootgenshort}{network services\xspace}
  \newcommand{\brootgenlong}{network services like DNS\xspace}
  
  \newcommand{\merit}{the Merit darknet\xspace}
  \newcommand{\meritgen}{darknet\xspace}
  \newcommand{\meritgens}{darknets\xspace}
  \makeatletter
    \renewcommand\section{\@startsection{section}{1}{\z@}%
      {-.25\baselineskip \@plus -2\p@ \@minus -.2\p@}%
      {.1\baselineskip}%
      {\@secfont}}
    \renewcommand\subsection{\@startsection{subsection}{2}{\z@}%
      {-.15\baselineskip \@plus -2\p@ \@minus -.2\p@}%
      {.05\baselineskip}%
      {\@subsecfont}}
    \renewcommand\subsubsection{\@startsection{subsubsection}{3}{\z@}%
      {-.15\baselineskip \@plus -2\p@ \@minus -.2\p@}%
      {-2\p@}%
      {\@subsubsecfont\@adddotafter}}
   \makeatother
  \newcommand{\RelaxFloats}{
  	\renewcommand{\topfraction}{0.9}
  	\renewcommand{\floatpagefraction}{0.9}
  	\renewcommand{\textfraction}{0.1}
  }
  \RelaxFloats

\begin{document}
\pagestyle{plain}
\title[Durbin: Internet Outage Detection with Adaptive Passive Analysis]{Durbin: Internet Outage Detection \\ with Adaptive Passive Analysis}

%
  \author{Asma Enayet}
  \affiliation {
    \institution{USC/ISI and}
    \department{the Thomas Lord CS Dept.}
    \city{Los Angeles}
    \country{USA}
  }
  \author{John Heidemann}
  \affiliation {
    \institution{USC/ISI and}
    \department{the Thomas Lord CS Dept.}
    \city{Los Angeles}
    \country{USA}
  }
%
%
%
%
\begin{abstract}
Measuring Internet outages is important to allow ISPs to improve their services,
  users to choose providers by reliability,
  and governments to understand the reliability of their infrastructure.
Today's active outage detection
  provides good accuracy with tight temporal and spatial precision
  (around 10 minutes and IPv4 /24 blocks),
  but cannot see behind firewalls or
  into IPv6.
Systems using passive methods can see behind firewalls,
  but usually, relax spatial or temporal precision,
  reporting on whole countries or ASes at 5\,minute precision,
  or /24 IPv4 blocks with 25\,minute precision.
We propose Durbin, a new approach to passive outage detection that
  \emph{adapts spatial and temporal precision} to each network they study,
  thus providing good accuracy and wide coverage
  with the best possible spatial and temporal precision.
Durbin observes data from Internet services or network telescopes.
Durbin studies /24 blocks to provide fine spatial precision,
  and we show it provides good accuracy even for short outages (5 minutes)
  in 600k blocks with frequent data sources.
To retain accuracy for the 400k blocks with less activity,
  Durbin uses a coarser temporal precision of 25~minutes.
Including short outages is important:
  omitting short outages underestimates overall outage duration by 15\%,
  because
  5\% of all blocks have at least one short outage.
Finally, passive data allows Durbin to report this 
  results for outage detection in IPv6 for 15k /48 blocks.
Durbin's use of per-block adaptivity is the key to providing good accuracy and broad coverage
  across a diverse Internet.
\end{abstract}
\maketitle
\pagestyle{plain}

\section{Introduction}

Internet outages are an economic and societal challenge.
An outage  costs Amazon  \$66k %
  on 2013-08-19~\cite{amazon_cost}. 
Data-centers lose \$5k per minute when users can not reach them~\cite{datacenter_cost}.
Natural disasters, political events, software and hardware failure, human error,
  and malicious activity can cause Internet outages~\cite{comprehensive, bogutz2019identifying, quan2014internet, dainotti2011analysis}.
\reviewfix{}
Prior outage detection has shown outages are rare but  ubiquitous~\cite{song2004internet,dainotti2011analysis,quan2012detecting,padmanabhan2019find,enayet2022internet}.
Monitoring approaches use active measurement or passive traffic analysis.
Active monitoring has vantage points (VPs) query destinations,
  responders prove reachability~\cite{katz2008hubble,schulman2011pingin,katz2012lifeguard,quan2013trinocular, padmanabhan2018analyzing}.
Passive methods instead infer outages by the absence of prior network traffic~\cite{shah2017disco,richter2018advancing,guillot2019chocolatine,kirci2022my}.

Today's active outage detection
  provides good accuracy with tight temporal and spatial precision
  (around 10 minutes and IPv4 /24 network blocks),
  but they face two limitations.
First, traffic from active observations draws abuse complaints and blocking from those who consider it intrusive.
Second, active methods cannot see behind firewalls.

Passive outage detection
  today usually relaxes spatial or temporal precision,
  reporting on whole countries or ASes at 5\,minute precision~\cite{guillot2019chocolatine},
  or /24 IPv4 blocks with 30\,minute precision~\cite{richter2018advancing},
  or require physical devices and so have limited coverage~\cite{shah2017disco}.
However, passive systems have some advantages:
  they pose no additional traffic on targets
  and so do not draw abuse complaints or blocking.
In addition, they can see networks behind firewalls
  when those networks send traffic.
\reviewfix{}
This paper proposes Durbin, a system
  that detects Internet outages based on passive analysis of data sources.
Unlike prior passive systems,
  Durbin's detection algorithm is parameterized based on the historical data of each block, allowing it to cover both /24 IPv4 blocks and extend coverage to /48 IPv6 blocks.
Durbin can vary the spatial and temporal precision of detection which provides high accuracy, even in networks with weaker signals, enhancing Durbin's effectiveness across a wider range of network conditions.
Our approach provides a new, systematic approach to passive analysis to address these problems, making the following three contributions:

\textbf{Detecting short outages: }
\reviewfix{}
We know brief outages occur 
(short-burst DDoS attacks or pulse attacks, but prior systems do not detect outages shorter than 10 minutes for individual /24 blocks.
Prior active detection systems use active probing and probes every 11 minutes,
  and cannot increase temporal precision without becoming excessively intrusive
  (which could result in abuse complaints or silent discard of measurements).
Prior passive detection systems detect short-duration outages,
  but at the cost of providing only much coarser, AS-level spatial precision.

Our new approach interprets passive data
  and can employ exact timestamps of observed data,
  allowing both fine spatial and temporal precision when possible.
Our measurements show in \autoref{sec:validation_shortoutages} that around 5\% of total blocks have 5\,minute outages that were
  not seen in prior work.
These short outages add up---when we add the outages from 5 to 10\,minutes that were previously omitted to observations,
  we see that total outage duration increases by 20\%.

\textbf{Optimizing across a diverse Internet:}
Variation in Internet use means that outage detection systems should 
  be tuned to operate well in each region.
Our second contribution is to describe the first passive system that optimizes parameters for each block
  to provide fine spatial and temporal precision when possible,
  but can fall back to coarser temporal precision when necessary.
\reviewfix{}
By contrast,
  although prior passive systems optimize some parameters,
  they operate with a homogeneous global sensitivity,
  and therefore provide only
  coarse spatial coverage (at the country or AS level~\cite{guillot2019chocolatine},
  or decreasing coverage).
We instead exploit the ability to trade-off between spatial and temporal precision (\autoref{sec:validation_trade-off}),
  allowing some blocks to have less temporal precision.
This flexibility means we can retain accuracy
  and increase coverage by reporting coarser results for blocks that otherwise
  would be ignored as unmeasurable.
Our hybrid approach detects around 20\% more outage duration than fixed parameters (\autoref{sec:validation_trade-off}).

\textbf{Extending to IPv6:}
Our third contribution is to show that our new approach applies
  to IPv6 (\autoref{sec:results}).
Since passive data comes from live networks,
  we avoid the active methods requirement for an accurate IPv6 hitlist.
\reviewfix{} 
Although there has been considerable work on IPv6 hitlists~\cite{beverly2015measuring, gasser2018clusters,Foremski16a, Murdock17a, song2022det, Beverly18a},
  their IPv6 coverage remains incomplete,
  particularly in the face of client preference for privacy-preserving addresses
  that are intentionally hard to discover.
\reviewfix{}
We show that coverage using passive data from \broot and
  an example of \brootgenlong,
  is about 17\% of the Gasser hitlist~\cite{gasser2018clusters},
  and our approach could cover 5$\times$  more than Gasser
  if it used data from Wikipedia or $10^5\times$ more if using NTP~\cite{rye2023ipv6}.

\textbf{Data availability and ethics:}
Our work uses publicly available datasets.
\PostSubmission{}
Datasets for the input and results from our experiments are available at
  no charge~\cite{passive_outage_data}.
Our analysis uses data from services, not individuals, so it poses no privacy concerns.
We discuss research ethics in detail in \autoref{sec:research_ethics}.

\section{Related Work}
	\label{sec:related}

Internet outage detection systems can use either active data monitoring or passive data monitoring. 
We highlight how the outage detection system we propose compares to existing active and passive monitoring techniques for both IPv4 and IPv6 address blocks.

\textbf{Outage Detection Systems using Active Monitoring}
  probe the Internet from a set of vantage points,
  typically distributed across different networks,
  sending pings or traceroutes to most or all of the Internet.

ThunderPing first used active measurement to track weather-related outages~\cite{schulman2011pingin}.
They probe many individual addresses in areas with severe weather from around ten vantage points and report outages for individual addresses.
\reviewfix{}
Padmanabhan et al.~later showed that outages sometimes occur at spatial
  scales smaller than a /24 block~\cite{padmanabhan2019find}.
Like this prior work, we are interested in edge networks,
  but our goal is wide coverage, not just areas under severe weather.

Hubble finds potential Internet outages by probing all~.1 addresses,%
  triggering traceroutes to %
  localize a potential outage~\cite{katz2008hubble}. 
LIFEGUARD extends Hubble to work around local outages caused by routing~\cite{katz2012lifeguard},
  detecting outages per routable prefix.
We instead do passive traffic observations from network-wide services,
\reviewfix{} and target outages in edge networks,
  using finer IPv4 /24s and also adding IPv6.

Trinocular provides precise measurements of Internet reliability to all ``measurable'' edge networks, 5.1 million /24 blocks in current datasets~\cite{quan2013trinocular}.
It uses adaptive ICMP probing every 11\,minutes
  dynamically adapting how many probes are sent to balance traffic and accuracy. 
Our passive work adds coverage for firewalled regions and extends to IPv6.

Our passive monitoring technique
  does not increase traffic on target networks, unlike active systems.
It also supports outage detection behind firewalls
  and we are the first to report results for IPv6.
Finally, we support finer temporal resolution (up to 5\,minutes)
  when possible,
  while retaining fine spatial resolution (IPv4 /24s).

\textbf{Outage Detection Systems using Passive Monitoring}
  infer outages by the disappearance of previously observed traffic.
Passive systems watch traffic from some global network service such as
  a CDN, a DNS service, or Internet background radiation seen in
  network telescopes. 

Dainotti et al.~examined outages from censorship~\cite{dainotti2011analysis}
  as detected in observations from both BGP and
  Internet background radiation (IBR) as seen in
  network telescope traffic~\cite{moore2004networktelescope}.
Chocolatine~\cite{guillot2019chocolatine}
  formalizes this approach, uses SARIMA models to detect outages in IBR~\cite{pang2004IBR}.
\reviewfix{}
Each of these passive systems improves temporal resolution,
  either by using more input traffic or by growing spatial precision.
For example, Chocolatine has 5\,minute temporal precision,
  but only at the scale of entire ASes.
We instead infer outages using data from passive sources which provide important insight into networks that block active probes.

CDN-based analysis provides /24
  spatial precision at 1\,hour temporal precision~\cite{richter2018advancing}.
As a passive approach, it also provides global results
  for firewalled blocks,
  with
  broad coverage (2M /24 blocks).
It optimizes expected response from the history of each block,
  but sets overall detection parameters globally.
We provide good spatial \emph{and} temporal precision,
  in part with additional per-block customization,
  and our approach could apply to CDN data
  to similar coverage.

Fontugne et al.~use RIPE Atlas data to
  identify network outages~\cite{staff2015ripe}.
Blink detects failures without controller interaction by analyzing TCP retransmissions~\cite{holterbach2019blink}. 
Blink creates a characteristic failure
signal when multiple flows aggregate retransmission information.

Our proposal differs from prior passive systems for several reasons.
First, protocols such as DNS or NTP can provide much larger coverage than RIPE:
  we infer outages from existing networks instead of explicit status provided by the network service providers.
Second, we can ensure finer temporal precision (5 minutes or less) by interpreting passive data and using exact timestamps of observed data. 
We optimize parameters for each block to provide precise results but may use coarser temporal precision when required.
We discuss precision in detail in \autoref{sec:sensitivity_timebin} and \autoref{sec:validation_trade-off}.

\textbf{Hybrid Active and Passive Detection Systems:}
Disco detects outages by passively detecting correlated disconnection events
  across actively maintained connections from about 10k sites~\cite{shah2017disco}.
We instead detect outages by analyzing passive traffic with IP and timestamps and search for a gap in that traffic without injecting excess traffic.
Our system customizes parameters for each block to optimize the performance of our model.

\textbf{IPv6 Coverage:}
Prior work on outage detection only considers IPv4.
IPv6 hitlists~\cite{Gasser16a, Foremski16a,Murdock17a, gasser2018clusters, Beverly18a, song2022det}
  are a potential step towards IPv6 outage detection.
Gasser et al. reported the first large IPv6 hitlist~\cite{Gasser16a}
  using prior data %
  and traceroutes to
  find 25.5k prefixes, 21\% of what was announced at the time.

Building on this work,
Beverly et al.~use random probing
  to discover 1.3M IPv6 routers~\cite{Beverly18a}.
Entropy/IP models IPv6 use with information-theoretic and machine-learning techniques~\cite{Foremski16a}.
After training on 1k addresses,
  40\% of their 1M candidate addresses are active.
Murdock et al.~search regions near known %
  addresses,
  discovering 55\,M new active addresses~\cite{Murdock17a}.
AddrMiner also grows the hitlist from a seed set~\cite{song2022addrminer},
  finding 1.7\,B addresses after dealiasing. %
Rye and Levin instead turn to passive addresses in NTP traffic,
  finding an impressive 7.9\,B addresses~\cite{rye2023ipv6},
  showing that prior approaches missed many client addresses.

We directly use passive data, like Rye and Levin,
  avoiding the need to build a hitlist for active probing.
We provide the first reports of outage detection in IPv6.

\section{Methodology}
  \label{sec:methodology}
  \label{sec:durbin_algorithm}

Durbin methodology is to
  observe passive data from some service  (\autoref{sec:data_requirements}).
From history,
  it models the probability traffic arrives from any address
  in some time period (\autoref{sec:durbin_history}).
It then detect deviations from this model
  with Bayesian inference, reporting traffic reduction as an outage
  (\autoref{sec:durbin_address_level}).
Finally, when possible, it combines observations for multiple observers
  in a block (\autoref{sec:durbin_block_level}).
We optimize parameters 
  to trade off spatial and temporal precision
  for each block (\autoref{sec:parameters}).
  
\subsection{Data Requirements}
\label{sec:data_requirements}

Durbin uses passive traffic observations from
  \brootgenlong
  or \meritgens.
Many \brootgenshort could serve as input to Durbin:
  large websites like Google, Amazon, or Wikipedia;
  web infrastructure like CDNs;
  infrastructure services like DNS or NTP\@.
\meritgens are an alternative,
  with darknets operated by CAIDA and Merit
  and smaller telescopes operated by many parties.
Durbin's requirement is that the data source
  accurately reports communication
  from a client IP address at some specific time.
Ideally, it provides such data  real-time or near-real-time.
Our approach works equally well for IPv4 and IPv6,
  as we show in \autoref{sec:results}.

Although we can use many possible data sources,
we evaluate our system using two specific systems:
First, we use traffic arriving at 
  \broot~\cite{broot},
  one of the 13 authoritative DNS Root services~\cite{RootServers16a}.
Second, we evaluate it using passive traffic arriving
  at  \merit~\cite{Merit21a}.
These two very different data sources
  show that Durbin generalizes 
  and can apply to many potential passive data sources.
  
Each data source
  collects traffic
  and shares the time and partially-anonymized source IP address of each flow.
We take several steps to minimize any privacy risks to users generating
  the data.
For \broot we omit other fields (such as query name)
  that are not required for detection.
For \merit we retain most fields only until we filter for spoofing.
Durbin models traffic at the block level,
  so we preserve the network portion of the IP address (IPv4: /24, IPv6: /48)
  and anonymize the remaining bits
  to shuffle individual users.

We see \broot as representative of a large Internet service
  that receives global traffic.
For
  \broot,
  we see about
  700M queries from about 7M unique locations per day.
While large, this coverage is much smaller
  then large websites like Google, Amazon, or Wikipedia.
In \autoref{sec:IPv6coverage} we quantify \broot coverage,
  show that Wikipedia would provide better coverage than today's public IPv6 hitlists.
Thus the smaller coverage of \broot represents a limitation
  in the data that we have access to,
  but \emph{not} a fundamental limitation
  in our approach.
Although \broot does receive spoofed traffic when it is attacked,
  when not under attack all queries arriving at \broot
  are from legitimate recursive resolvers,
  and we assume the source address indicates a valid, active IP address.
We believe the Durbin algorithm can apply to other data sources
  (such as NTP~\cite{rye2023ipv6}),
  although any new source will require tuning
  Durbin parameters, as we do for our current sources (\autoref{sec:parameters}).

We also evaluate over \merit, an example \meritgen.
A darknet does not run real services
  and so should receive no legitimate traffic.
Incoming traffic is often %
  network scanners,
  malware attempting to propagate itself,
  or backscatter, where someone spoofed the darknet as the source IP
    for traffic sent to another party.
A darknet's source addresses suggest a live network,
  but could be spoofed.
We filter darknet traffic to discard traffic where we do not believe
  the source address is legitimate.
We follow CAIDA's filtering rules~\cite{dainotti2013estimating},
  discarding
  packets with TTL exceeding 200 which are not ICMP;
  those with IPv4 sources ending in .0 or .255
  or identical source and destinations;
  and protocols 0 and 150.

\PostSubmission{
\reviewfix{I23C6,I23C13,I23C14,I23D1, I23D4: We also use RIPE Atlas data for short outage comparison.
Our system would change based on different input types when we use different possible data sources. 
For future work, we will add Merit darknet data as an input to our system and compare the results}}

\subsection{Learning From History}
\label{sec:durbin_history}

Durbin models
  expected traffic from address $a$ from long-term observations.
Each address has three parameters:
  the timebin duration for detection, $T(a)$;
  the historical probability we see traffic in that timebin, $\pi(a)$,
  and model has enough data to be consistent or \emph{measurable}, $M(a)$.
We discuss how  we
  generalize from addresses to blocks in \autoref{sec:durbin_block_level},
  and how set these parameters in
  \autoref{sec:parameters}.

We divide the timeline into specific timebins for each address,
each lasting $T(a)$ seconds. 
In principle, $T(a)$ may range from 1 to 60\,minutes,
  but currently we select from short and long options
  (typically 5 or 25\,minutes, \autoref{sec:parameters}).
\PostSubmission{}

The \emph{active probability} $\pi(a)$,
  is the probability that traffic arrives in timebin $T(a)$.
We compute $\pi(a)$ for each address
  from from long-term observation of address $a$,
  based on the last $d$ days.
We currently use $d$ of two days.

Finally, we define measurability %
  $M(a)$,
  when the address has enough data to provide signal,
  when $\pi(a) > \theta_{\Vs{measurable}}$.

\reviewfix{}
Durbin works well for short-term outages,
  but outages lasting longer than the training period
  will disappear when training considers only the outage.
Long-term outages are difficult to handle in most outage detection systems.
In general, external information is required to distinguish
  long outages from changes in usage.

\subsection{Address-Level Outage Detection}
	\label{sec:durbin_address_level}

To detect outages 
  we estimate the
  belief $B(a)$ as probability
  that address $a$ is reachable,
  from  0.0 to 1.0, certainly down to certainly up.
We classify an address as unreachable (down) when
  $B(a) < \theta_a$ and  reachable (up) when $B(a) > 0.95$.
We consider middle values ($\theta_a < B(a) < 0.95$)
  to indicate uncertainty,
  with the hope that information in the next timebin will resolve its state.
We set the threshold \emph{$\theta_a$} to 0.6,
  and validate each of these parameters 
  in \autoref{sec:sensitivity_beliefthreshold}.

\begin{table}
\begin{center}
\resizebox{0.8\columnwidth}{!}{
\begin{tabular}{ c|c|c } 
 \textbf{Observation} & \textbf{Prior} & \textbf{(Observation | Prior)} \\
 \hline
 Negative & Down & $1$ \\ 
 Negative & Up & $1-\pi(a)$ \\
 Positive & Down & $\pi(a)$ \\
 Positive & Up & $1- P(\R{neg}|\R{down})$
\end{tabular}
}
\end{center}
\caption{Responses when a timebin has traffic or not}
\label{tab:inferable_bayes}
\end{table}

\autoref{tab:inferable_bayes} shows how belief changes according to conditional probabilities.
We compute belief $B(a)$
  by applying Bayesian inference
  on a stream of observations in each timebin $T(a)$
  if the address has traffic (positive) or not (negative).
For each timebin with a positive and negative observations
  compute new belief ${B'}(a)$ from prior belief $B(a)$ as:
\begin{equation}
\label{eq:belief_up}
 {B'}(a) = \frac{\pi(a) B(a)}{\pi(a) B(a) + (1-P(no|down))(1- B(a))}
\end{equation}

\begin{equation}
\label{eq:belief_down}
 {B'}(a)  = \frac{(1-\pi(a)) B(a)}{(1-\pi(a)) B(a) + (P(no|down))(1- B(a))}
\end{equation}

We illustrate shifts in belief
  with case studies in \autoref{sec:case_studies}.
These equations get stuck when $B(a)$ reaches 0 or 1,
  so we limiting $B(a)$ to the range  $B_{\Vs{min}}$ to $B_{\Vs{max}}$,
  currently set to 0.1 and 0.95.

\subsection{Block-Level Outage Detection}
\label{sec:durbin_block_level}

We next merge address-level belief
  ($B(a)$) of all addresses in a block to determine block-level belief $B(b)$.

We study /24 address blocks as the smallest unit of spatial coverage,
  following prior work~\cite{quan2013trinocular, richter2018advancing}.
Combining results from multiple addresses
  in one block can improve accuracy since we can get more information about the block.

Following our definition of addresses,
  we consider block status
  in timebins of duration $T(b)$,
  and that value can vary by block.
This approach follows address analysis with $T(a)$ in \autoref{sec:durbin_history}.
We define $B(b)$ as the belief in the status of each /24 block $b$.
Unlike addresses $B(b)$ is not inferred from data,
  but instead, it combines all address beliefs in that box,
  defining:
\[
B(b) = \max(B(a_i)) \forall{a_i} \in b
\]

We merge address-level detection results for each $T(b)$ to get block-level results.
Block detection uses a potentially different threshold ($\theta_b$),
  identifying an outage when $B(b) < \theta_b$,
  a reachable block when $B(b) > 0.95$,
  and otherwise identifying the block as uncertain.
We set $\theta_b = \max_{a \in b}\theta_a$.

When considering block-level outage detection
  we  select other block-level parameters from the address-specific parameters,
  taking the minimum timebin and 
  considering a block measurable $M(b)$ if any address is measurable:
\[
T(b) = \min_{a \in b}T(a)  \quad \textrm{and} \quad
M(b) = \bigvee_{a \in b}M(a)
\]

%
%
%
%
%
%
%
%
%
%
%
%

%

\subsection{Optimizing Parameters for Each Block}
	\label{sec:parameters}

We optimize Durbin parameters for each block
  to trade-off accuracy and coverage.
We adjust timebin duration ($T(b)$)
  and thresholds for measurability ($\theta_b$,  $\theta_{\Vs{sparse}}$ and $\theta_{\Vs{measurable}}$)
  based on amount of traffic to each block.

\comment{}
Durbin first selects timebin duration ($T(a)$)
  to provide rapid detection for addresses with frequent traffic and reliable detection
  for addresses with sparse traffic.
Currently we select between two timebin durations,
  based on the threshold $\theta_{\Vs{sparse}}$,
  labeling addresses where $\theta_{\Vs{measurable}} \le \pi(a) < \theta_{\Vs{sparse}}$ as sparse
  and those with $pi(a) > \theta_{\Vs{sparse}}$ as frequent.
With \broot as our data source
  we set $\theta_{\Vs{sparse}}$ as 0.6,
  and use 5 and 25\,minutes as short and long timebins.
With \merit as the source we
  set $\theta_{\Vs{measurable}}$
  and  $\theta_{\Vs{sparse}}$ as 0.6
  and $T(b)$ is 20\,minutes.
We use the same Durbin algorithm
  for each source,
  but choose parameters ($\theta_{\Vs{sparse}}$ and $T(\cdot)$)
  based on 
  analysis of coverage and accuracy in \autoref{sec:sensitivity_timebin} and \autoref{sec:validation_trade-off} as we vary $T(a)$.
As future work, we plan to vary $T(b)$ for merit
  and to explore allowing $T(a)$ to vary
  continuously.  %

Finally, we define $T(b) = \min(T(a_i)), \forall a_i \in b$:
  the block can be as sensitive as the best address,
  analogous to belief following the most reliable address
  (\autoref{sec:durbin_block_level}).

Currently thresholds %
  ($\theta_a$ and $\theta_b$),
  are fixed (both at 0.6).
Because belief ranges from 0 to 1, fixed values here seem appropriate.
However, belief adapts based on block history (\autoref{sec:durbin_address_level})
  and so it is customized for each block.

\section{Demonstrating Viability}
	\label{sec:case_studies}

We next show that Durbin feasible,
  first by confirming that a network-wide service
  has enough data to provide coverage of many blocks,
  then by showing how Durbin handles sources with different amounts of traffic.

\subsection{Traffic Rates per Address}
    \label{sec:traffic_per_address}

We first characterize traffic
  in our data sources. %

\begin{figure}
\centering
\includegraphics[width=0.86\linewidth]{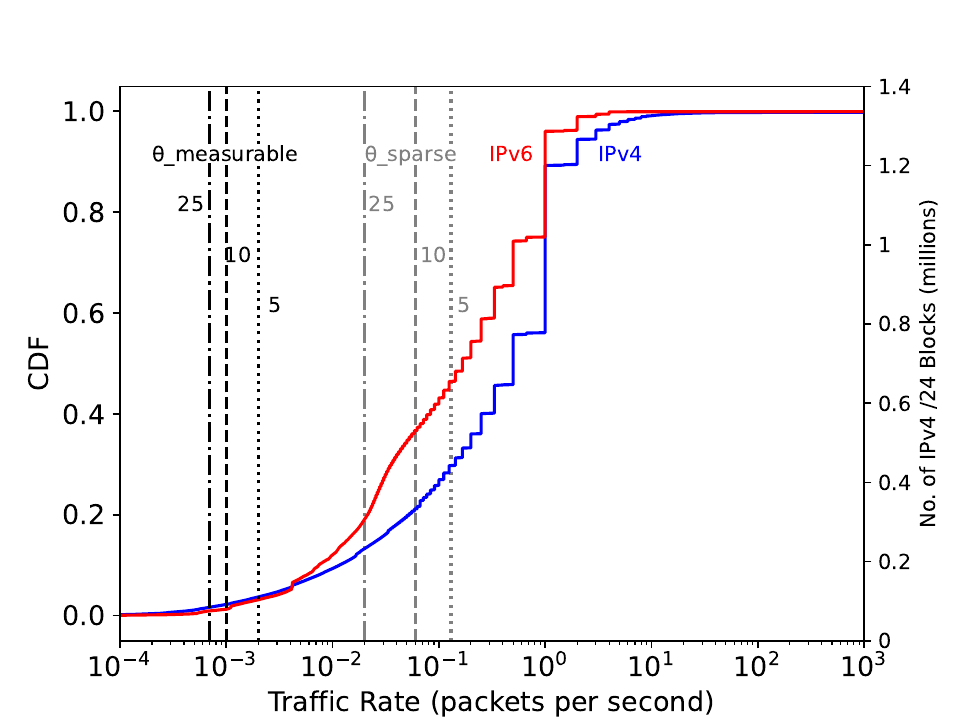}
\caption{Traffic per address for \broot on 2019-01-10}
\label{fig:rates_per_address}
\end{figure}

\subsubsection{Traffic Distribution in \broot}
\autoref{fig:rates_per_address} shows a cumulative
  distribution of traffic arrival rates
  for each address in \broot
  by IPv4 (the blue line) and IPv6 (red) over one day.

Both IPv4 and IPv6 show similar distributions,
  although IPv4 reports on 1.2M blocks while IPv6 is only 13k.
The similarity of these distributions suggests
  the same algorithm will apply to both IPv4 and IPv6.
This similarity is important because we show good accuracy for IPv4
  (\autoref{sec:validation_long-term}) where we can compare to alternative methods.
As the first public IPv6 approach,
  we cannot compare to alternatives there,
  but since it is the same algorithm, we expect IPv4 accuracy to
  apply to IPv6 as well.

This data allows us to estimate coverage of Durbin-with-\broot.
Coverage depends on our parameters 
  for timebin ($T(b)$, \autoref{sec:parameters})
  and measurability ($\theta_{\Vs{measurable}}$, \autoref{sec:durbin_history}),
    as we evaluate in \autoref{sec:sensitivity_timebin}.
Here we see measurability for three different timebin sizes
  (25, 10, and 5\,minutes, vertical black lines from left to right).
Almost all \broot sources are measurable (90\% for 5\,minutes, and 95\% for 25\,minutes).

The gray vertical lines show the division between sparse and frequent sources,
  determined by $\theta_{\Vs{sparse}}$.
Most sources (80\% of IPv4 and 50\% of IPv6) and have frequent data
  (to the right of the gray vertical line).
The ability to pick up sparse sources
  (blocks between the black and gray lines)
  adds coverage for another 20\% or 50\% of all blocks (100k to 400k for IPv4, and 1000 to 6500  for IPv6).
  
\subsubsection{Traffic Distribution in \merit}

Traffic in \merit differs
  because it is not an active service,
  but receives only unsolicited  traffic.
Similar analyses reveal a wider variability in traffic arrival probabilities for IPv4.

\PostSubmission{}
The distribution of traffic in \merit is similar to that of \broot,
  suggesting that the same algorithms apply, although perhaps with different parameters.
However, \merit has many more sparse blocks (80\%, compared to 30\% for \broot).
The smaller amount of traffic in a darknet shows the importance of observing
  network services,
  but scanning detected in darknets can
  can provide information about the status of otherwise
  silent and firewalled blocks.

\subsection{Belief for Frequent Traffic}
\label{sec:belief_change_dense}

We next show how Durbin can react very quickly when
  an address has frequent traffic.
From \autoref{sec:durbin_address_level}, an address $a$ has frequent traffic if $\pi(a) > \theta_{\Vs{sparse}} $
Here we pick one example address where  $\pi(a)$ is 0.9;
  we see similar results in the hundreds of other addresses
  that have similar traffic levels.

\begin{figure*}
\begin{center}
  \begin{subfigure}[t]{0.3\textwidth}
    \begin{subfigure}[t]{\textwidth}
      \includegraphics[width=\textwidth]{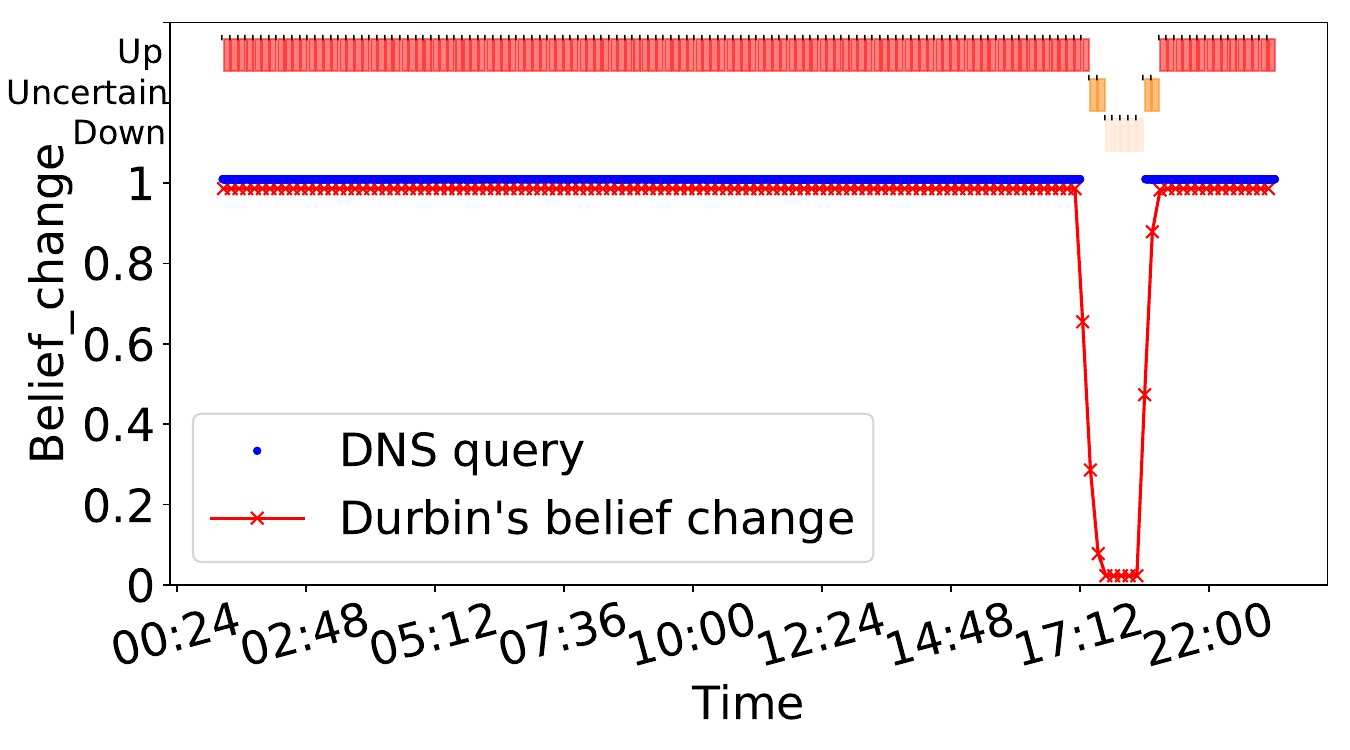}
      \caption{Frequent}
      \label{fig:dense_belief}
    \end{subfigure}
    \vspace{1em} %
    \begin{subfigure}[t]{\textwidth}
      \includegraphics[width=\textwidth]{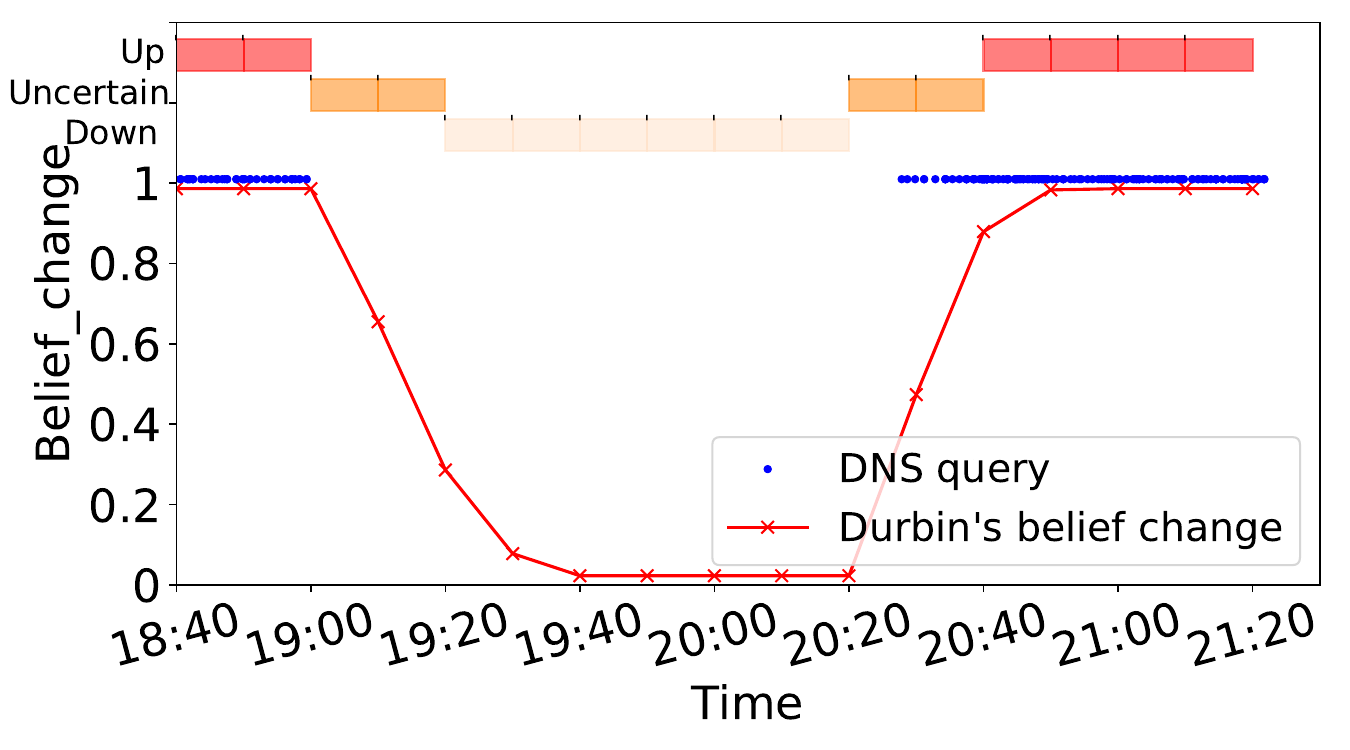}
      \caption{Zoomed (frequent) }
      \label{fig:dense_belief_zoomed}
    \end{subfigure}
  \end{subfigure}
  \quad
  \begin{subfigure}[t]{0.3\textwidth}
    \begin{subfigure}[t]{\textwidth}
      \includegraphics[width=\textwidth]{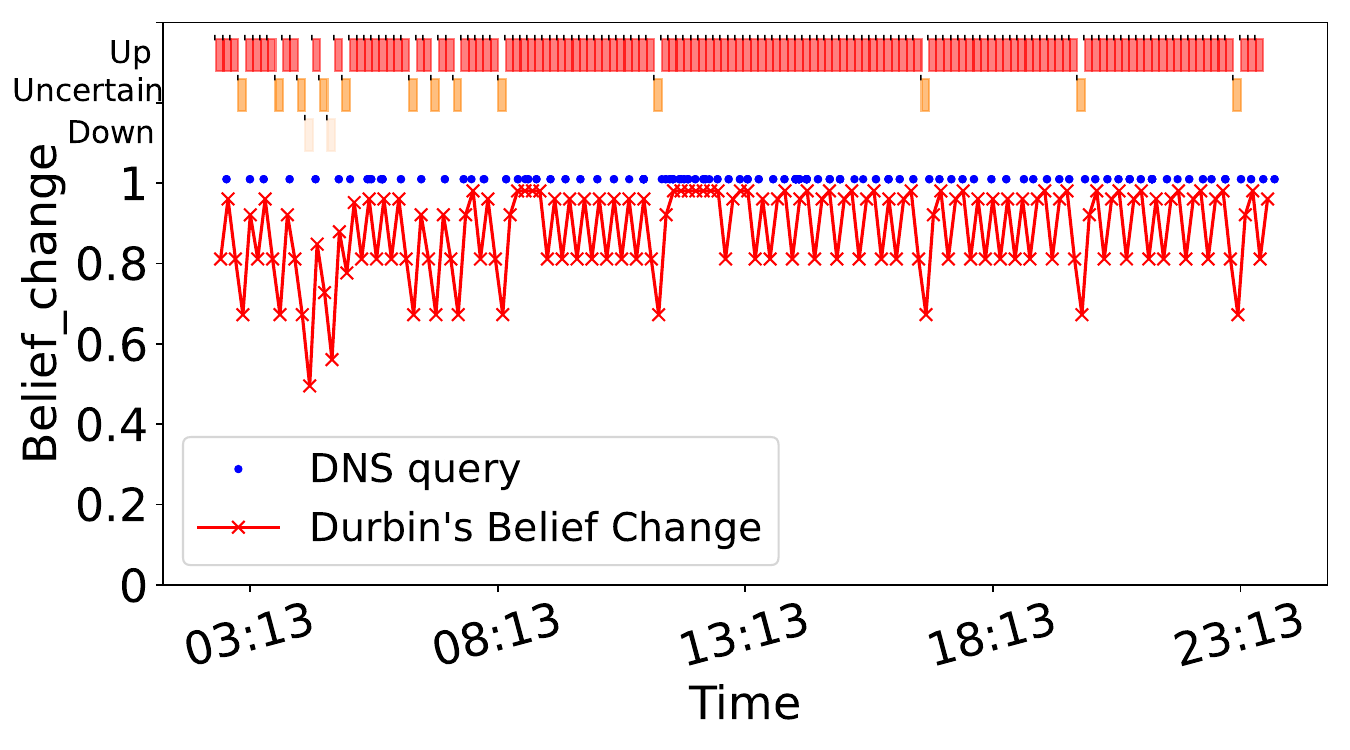}
      \caption{Sparse, 10\,minute timebin}
      \label{fig:sparse_belief}
    \end{subfigure}
    \vspace{1em} %
    \begin{subfigure}[t]{\textwidth}
      \includegraphics[width=\textwidth]{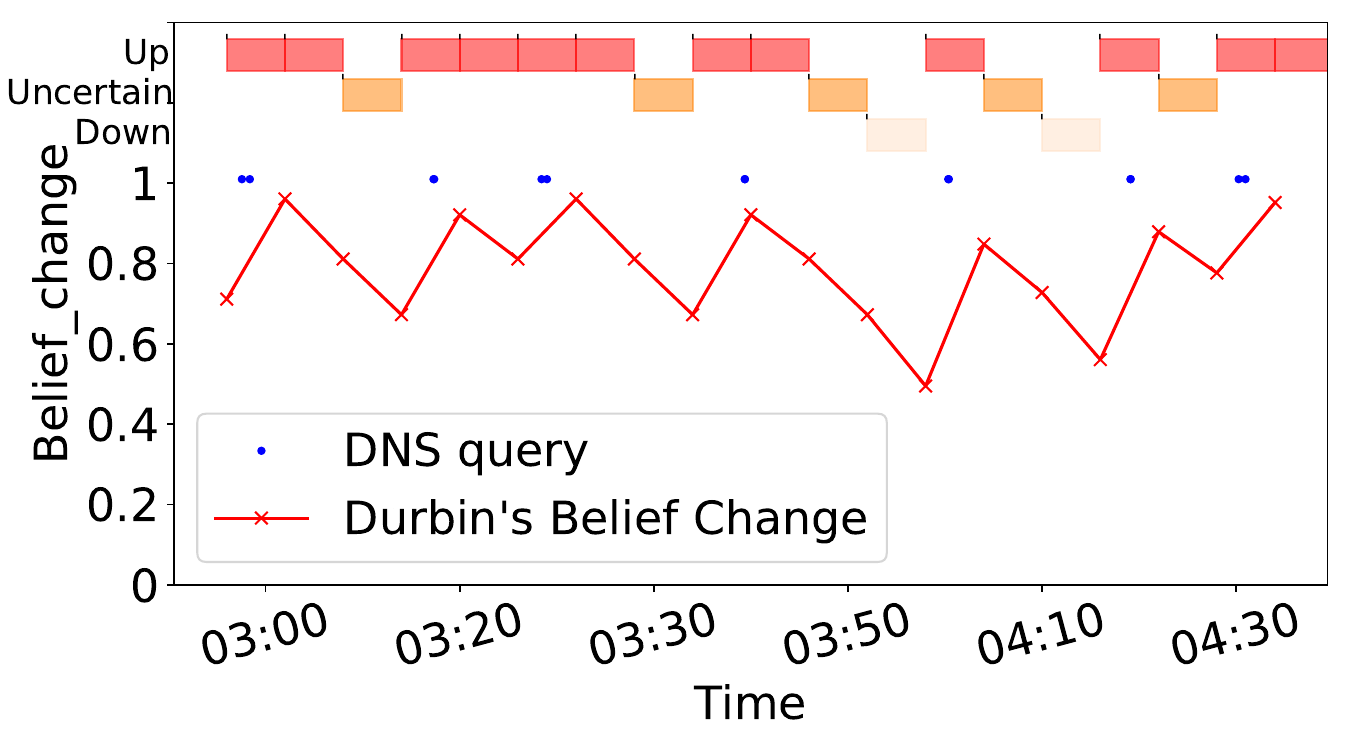}
      \caption{Zoomed sparse, 10\,minute timebin}
      \label{fig:sparse_belief_zoomed}
    \end{subfigure}
  \end{subfigure}
  \quad
  \begin{subfigure}[t]{0.3\textwidth}
    \begin{subfigure}[t]{\textwidth}
      \includegraphics[width=\textwidth]{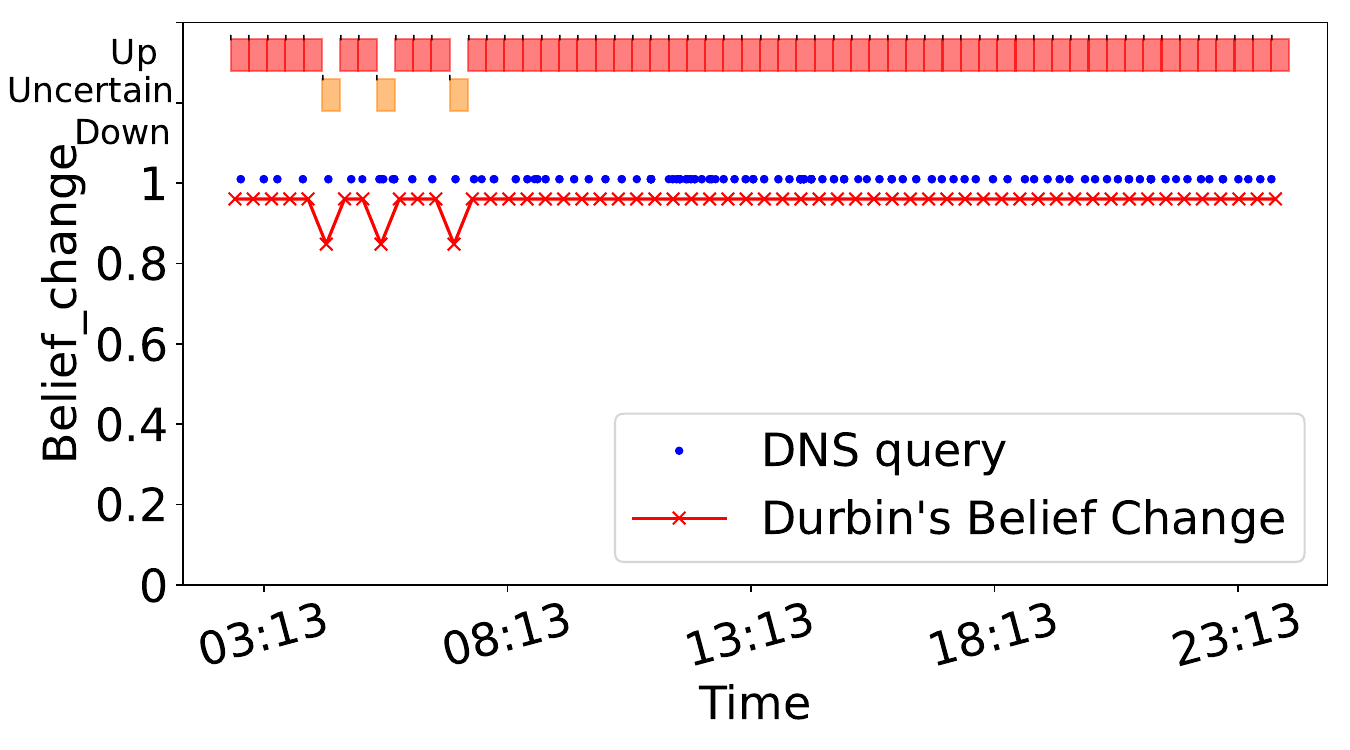}
      \caption{Sparse, 15\,minute timebin}
      \label{fig:sparse_belief_longbin}
    \end{subfigure}
    \vspace{1em} %
    \begin{subfigure}[t]{\textwidth}
      \includegraphics[width=\textwidth]{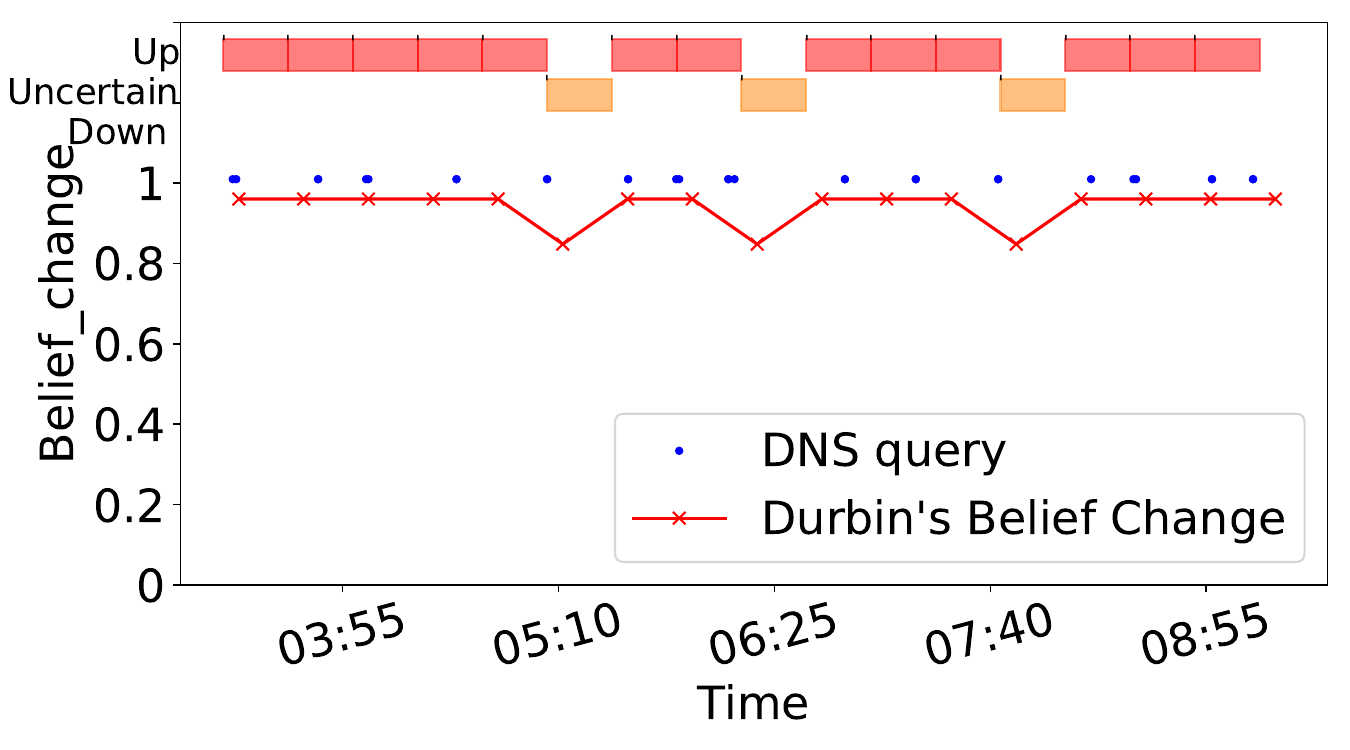}
      \caption{Zoomed sparse, 15\,minute timebin}
      \label{fig:sparse_belief_longbin_zoomed}
    \end{subfigure}
  \end{subfigure}
\end{center}
  \caption{Belief change in Durbin-with-\broot, varying traffic frequency and timebin duration. Data: 2019-01-10.}
  \label{fig:belief_change}
\end{figure*}


For addresses with frequent traffic,
  gaps are very unusual,
  so belief changes quickly from certainly up to down.
\autoref{fig:dense_belief} shows belief (the red line)
  for our example addresses with frequent traffic (blue dots).
In \autoref{fig:dense_belief},
  blue dots indicate traffic to an address at \broot on 2019-01-12
  with a gap from 19:20 to 20:20.

We show timebins as boxes at the top of \autoref{fig:dense_belief},
  showing reachable with red boxes (top row, with traffic),
  uncertain as dark orange in the middle,
  and unreachable (no traffic) with light orange, on the third row.
We then zoom on the outage period in \autoref{fig:dense_belief_zoomed}.

\PostSubmission{looking at figure 2: is downward belief change the same as upwards?
Should it be?  Isn't getting a single packet through proof the network is back?
If not, say here why not.
ALTERNATIVELY, maybe this is what we do for now,
but maybe we need ``precision improvement'' for this algorithm,
where once we decide we're up, we back time up and correct prior assertions.
We can do the same for Down.
We can add that algorithm AFTER quals, but you might mention it as possible future work now.
(Note that this algorithm is similar to what Guillermo does with FBS. ---johnh 2023-05-12}

\PostSubmission{
Alternatively, we could consider implementing a "precision improvement" for the algorithm, which involves going back in time and correcting prior assertions once we decide that the network is up or down. 
By implementing this precision improvement, we can improve the accuracy of the network status measurements and ensure that the system is more effective in detecting network outages.
This modification is our potential future work.
}

\autoref{fig:dense_belief}
  shows that frequent traffic (the solid blue dots before 19:00)
  yield confident of reachability
  (belief at the 0.90 maximum).
However, 
  the traffic gap from 18:56 to 20:26 (see \autoref{fig:dense_belief_zoomed})
  causes
  belief to drop through uncertainty (two timebins at 19:00)
  to unreachable (at 19:20).
This example shows how Durbin reacts quickly given
  frequent traffic.

\subsection{Belief for Addresses with Sparse Traffic}
\label{sec:belief_change_sparse}

For addresses with sparse traffic (when $\pi(a) < \theta_{\Vs{sparse}}$),
  Durbin must be more cautious in determining when an outage occurs.
Here we pick one example with $\pi(a)=0.6$.
\PostSubmission{}
Again we see similar results in the hundreds of other, sparse
  addresses where $\theta_{\Vs{measurable}} < \pi(a) < \theta_{\Vs{sparse}}$.

\autoref{fig:sparse_belief} shows one day (2019-01-13)
  for this representative address,
  with sparser traffic (blue dots)
  causing belief (the red line) to vary.
To see Durbin coping with gaps in data,  
  \autoref{fig:sparse_belief_zoomed} zooms in a 90\,minute period starting at 2:50.
Here belief often drops as multiple timebins pass without
  traffic, all the way to the outage
  at 3:40 and  4:10.
While relatively rare, these false outages are a problem
  that comes from attempting to track blocks with sparse traffic---we simply
  do not have enough information in a short timebin to make consistently good decisions.
We can discard such blocks, but that reduces coverage.
In the next section, we show how to gather more information to correct
  this situation.
\subsection{Sparse Traffic with a Longer Timebin} %
\label{sec:belief_change_sparse_longbin}

The false outages in \autoref{fig:sparse_belief_zoomed}
  result from short timebins not observing enough information 
  to make good decisions.
We next show that a longer timebin addresses this problem.

\autoref{fig:sparse_belief_longbin} shows one day
  for the same address as \autoref{fig:sparse_belief}),
  now with  a 25\,minute timebin.

With longer timebins, $\pi(a)$ rises to 0.9, enough information that we never
  detect false outages, and we see only three unknown periods
  for the same day results in several false outages and many unknown periods
  with a shorter timebin.
\autoref{fig:sparse_belief_longbin_zoomed} zooms 5\,hours starting at 3:00
  to show how a longer timebin bridges gaps.

We use these examples to motivate our overall design choices:
using Bayesian Inference to adapt belief based on multiple rounds of
    observation,
  and adapting a belief threshold and tuning parameters based on address history (\autoref{sec:sensitivity_beliefthreshold}).

\section{Validation of our Approach}
\label{sec:validation}

We next validate our results,
  examining accuracy with
  positive predictive value (PPV),
  recall, and true negative rate,
  sensitivity to the parameters of belief threshold and timebin duration.
We use \emph{PPV}
  instead of information retrieval's \emph{precision}
  to avoid confusion with terms temporal and spatial precision,
  both key ``knobs'' in our design. 
We then demonstrate that our approach can detect short outages and trade-off spatial and temporal precision.

We validate our results by comparing  them to prior data sources.
We run Durbin with both 
  both 7 days of \broot and 7 days of darknet as data sources.
We compare to   
  Trinocular~\cite{quan2013trinocular},
  a system using active analysis with very broad coverage (about 5M IPv4 /24s).
We validate our short-duration outage results by comparing them to Disco~\cite{shah2017disco}.
It uses RIPE Atlas data to detect correlated disconnections,
  making it sensitive to short-duration outages.
We would like to compare to other passive systems,
  but Durbin's much greater spatial sensitivity
  makes direct comparison to Chocolatine impossible~\cite{guillot2019chocolatine},
  and CDN data is not publicly available~\cite{richter2018advancing}.

We show Durbin system has good accuracy (\autoref{sec:accuracy}), how belief threshold (\autoref{sec:sensitivity_beliefthreshold}) and time-bin (\autoref{sec:sensitivity_timebin}) affects accuracy and coverage, we can detect short outages (\autoref{sec:validation_shortoutages}), Durbin can trade-off between temporal and spatial precision (\autoref{sec:validation_trade-off}) and our system is consistent over a long period of time (\autoref{sec:validation_long-term}).

\PostSubmission{}

\PostSubmission{we now have plans to get all of 2023-01.  You should plan to run on this as soon as you can.
That could be post-quals, or it could be an update between quals submission and quals final.
---johnh 2023-04-20}

\subsection{Accuracy of Durbin-with-\brootgenshort}
\label{sec:accuracy}


\subsubsection{Direct comparison of Durbin-with-\broot} 
	\label{sec:accuracy_full}

To compare Durbin-with-\broot to Trinocular
  we first find all /24 blocks observed both,
  yielding about 880k blocks in both datasets
  for the 7 days starting on 2019-01-09.
We then compare the block-durations (in seconds)
  each system identifies as reachable or not.

\autoref{tab:accuracy_before_correction} shows this confusion matrix,
    defining Trinocular outages as ground truth.
We define a false outage (\V{fo}) for a block when
  Durbin predicts the block is unreachable, but Trinocular can reach it,
  and have similar definitions for false availability (\V{fa}), true availability (\V{ta}), and true outages (\V{to}).

\begin{table*}
\centering
\begin{minipage}{\textwidth}
\begin{subtable}{0.32\textwidth} %
\centering
\resizebox{\textwidth}{!}{ %
\begin{tabular}{ r|r }
   	&  seconds \\
 \hline
  \rowcolor[HTML]{a0c080}
 True availability = TP &  52,525,765,695  \\
 \rowcolor[HTML]{C5E1A5}
 True outage = TN &  13,147,965  \\
 \rowcolor[HTML]{FFCCBC}
 False availability = FP  & 2,471,178 \\
 \rowcolor[HTML]{FAD9D9}
 False outage = FN & 31,087,360,212 \\
 PPV & 0.9999\\
 Recall & 0.6282\\
 TNR & 0.84178\\
\end{tabular}
}
\caption{Direct comparison of duration}
\label{tab:accuracy_before_correction}
\end{subtable}
\quad
\begin{subtable}{0.32\textwidth} %

\centering
\resizebox{0.9 \textwidth}{!}{ %
\begin{tabular}{ r|r }
   	&  seconds \\
 \hline
   \rowcolor[HTML]{a0c080}
 True availability = TP &  52,525,765,695 \\
\rowcolor[HTML]{C5E1A5}
 True outage = TN & 13,147,965 \\
 \rowcolor[HTML]{FFCCBC}
 False availability = FP  & 2,471,178 \\
  \rowcolor[HTML]{FAD9D9}
 False outage = FN & 78,163,261 \\
 PPV & 0.9999\\
 Recall & 0.9985\\
 TNR & 0.8417\\
\end{tabular}
}
\caption{Precision-aware comparison of duration}
\label{tab:accuracy_after_correction_seconds}
\end{subtable}
\quad
\begin{subtable}{0.28\textwidth} %
\centering
\resizebox{0.9 \textwidth}{!}{ %
\begin{tabular}{ r|r }
   	&  events \\
 \hline
   \rowcolor[HTML]{a0c080}
   True availability = TP &  359,415 \\
\rowcolor[HTML]{C5E1A5}
 True outage = TN & 508 \\
 \rowcolor[HTML]{FFCCBC}
 False availability = FP  & 241 \\
  \rowcolor[HTML]{FAD9D9}
 False outage = FN & 24,218 \\
 PPV & 0.9976\\
 Recall & 0.9368\\
 TNR & 0.6782\\
\end{tabular}
}
\caption{Precision-aware comparison of events}
\label{tab:accuracy_after_correction_events}
\end{subtable}
\end{minipage}
\caption{Confusion matrix for long-duration outages for Durbin-with-\broot (Dataset: 2019q1)}
\label{tab:accuracy_long_duration}
\comment{}
\comment{}
\comment{}
\end{table*}

PPV is uniformly good ($(\V{ta}/(\V{ta} + \V{fa}))= 0.9999$):
  Durbin's reported availability is almost always correct.

We next consider
  true negative rate (TNR)
  to quantify what duration of outages we report are true ($(\V{to}/(\V{to} + \V{fa})$)\@.
Our TNR is good, at 0.8417, but lower than the PPV\@.
Strong TNR means we correctly estimate outage duration,
  but TNR is lower than PPV because outages are rare,
  making small differences between Durbin and Trinocular more noticeable.

\subsubsection{How does imprecise comparison affect accurate results?}
Both Trinocular and Durbin measure with fixed timebins,
  and misalignment between the two systems inevitably results in
  small differences
  (just like comparisons measured in whole numbers of meters and feet
  will never be identical).
Measurement precision results in lower-than-expected recall ( $(\V{ta}/(\V{ta} + \V{fo})) = 0.6282$).

Recall suggests that we often find shorter outages than Trinocular.
Trinocular's temporal precision
  is coarser than Durbin ($\pm 330$\,s vs. $\pm 150$\,s),
  so Durbin can detect shorter outages.
\PostSubmission{prior sen: that's a good idea, Durbin COULD use exact timestamps to target start and end.  BUT I think your implementation actually uses timebins, right?
It would be great to tighten precision, but until you do that, please make the text match what we do.  ---johnh 2023-05-03}  
\PostSubmission{}
\PostSubmission{}
\PostSubmission{}
Enhancing Durbin's precision using exact timestamps is a potential future research direction.

Temporal precision is affected by several factors.
Most important is choice of timebin duration, $T(b)$.
If we select precise timing with a small timebin,
  then we will either lose accuracy (because many timebins have no responses and so we need to use a lower belief threshold),
    or we must increase our spatial scale
    to provide reliable decisions.
Also, actual outages do not always line up with timebins,
  so any timebin-based system may report outages up to one timebin late.
\PostSubmission{}
Finally, depending on block history, it make take multiple timebins
  to shift belief.
\PostSubmission{}

\subsubsection{Re-evaluating with Precision-Aware Comparison:}
	\label{sec:accuracy_precision_aware}
Our measurement accuracy is determined by temporal precision,
  but smaller differences provide metrics that can be misleading,
  exaggerating differences that reflect random phase
  of measurements
  rather than differences in the underlying conclusions.
To factor out the measurement system and get at the underlying phenomena
  we next consider precision-aware comparison.

\reviewfix{}
We define \emph{precision-aware comparison} as
  ignore differences that are shorter than the measurement
  timebin for a given block.
Ignoring these short differences is justified
  timebin phase is arbitrary,
  and it reflects more on quantization of outage detection into timebins
  than on the actual correctness of the underlying method.
We keep any differences lasting longer then the block's timebin,
  since those represent real disagreement in results.
We then 
  computing PPV, recall and TNR on these ``precision-aware'' observations.
(Our precision-aware comparison  is analogous to how
  CDN-based outages were comparing to
  only Trinocular outages lasting longer one hour,
  the CDN quantum~\cite{richter2018advancing},
  however we present both the full data \autoref{sec:accuracy_full}
  and precision-aware comparisons here.)

\autoref{tab:accuracy_after_correction_seconds} shows comparisons with
  observations with precision-aware time bins.
Now, PPV is uniformly good (inference of blocks being reachable is nearly always correct) as before which is 0.9999.
Also, Recall rises to near-perfect 0.9985
  (from 0.6282),
  because alignment eliminates what would otherwise be many short, false outages.

The number and duration of false outage events drops to one-quarter of before,
  from 31.09\,Gs to 78.16\,Ms.
We believe these improved results better reflect the true ability of
  passive observation to detect events,
  once with rounded time bins.

\subsubsection{Comparison with event counts }
    \label{sec:event_counts}

Our prior comparisons consider block-seconds,
  giving longer differences heavier weight.
In \autoref{tab:accuracy_after_correction_events}
  we instead count events (state changes).

When considering events instead of time,
  recall is dominated by many blocks that never change state,
  and recall and TNR are disproportionately
  by a few blocks that frequently change state.
Since a stable block has one correct event, but a block that frequently changes
  state may have hundreds of changes,
  events magnify the effects of frequently changing blocks.
Thus events show lower recall and TNR.
As future work, we plan to look for these frequently changing blocks
  to account for them with more conservative parameters.

\subsubsection{IPv6 Correctness}
    \label{sec:ipv6_correctness}

We validate the Durbin algorithms above for IPv4,
  finding excellent PPV and good TNR.
For IPv4, we can validate against other systems,
  but there are no prior results for IPv6 against which to compare.

Fortunately, our \emph{IPv4 and IPv6 algorithms are identical},
  and we showed in \autoref{fig:rates_per_address} that
  many addresses have similar traffic rates.
Since correctness depends on traffic rate and regularity and these are similar
  in IPv4 and IPv6,
  we  expect our accuracy for IPv4 outage detection
  to apply to IPv6.

\subsection{Accuracy of Durbin-with-\meritgen}

To confirm Durbin's accuracy \merit, a different data source,
  we repeat these comparison.
  
\subsubsection{Directly comparing Durbin-with-\meritgen}

To compare Durbin using \merit data with Trinocular, we find all /24 blocks in both and compare the duration for each system labels as reachable or not.

We see 66,776 blocks in Durbin-with-\merit,
  compared to 5,210,923 blocks in Trinocular,
  with 59,979 blocks in the intersection.
\merit has 66,776 blocks,
    but 3,198 are only active because of spoofing.
    We, therefore, compare the 63,578 remaining to the 5,210,923 blocks in Trinocular,
    finding 59,979 in the intersection.

\begin{table*}
\centering
\begin{minipage}{0.9\textwidth} %
  \begin{subtable}{0.45\textwidth} %
    \centering %
    \resizebox{0.75\columnwidth}{!}{
    \begin{tabular}{ r|r }
      & seconds \\
      \hline
      \rowcolor[HTML]{a0c080}
      True availability = TP & 3,064,574,810 \\
      \rowcolor[HTML]{C5E1A5}
      True outage = TN & 9,833,309 \\
      \rowcolor[HTML]{FFCCBC}
      False availability = FP & 59,215,390 \\
      \rowcolor[HTML]{FAD9D9}
      False outage = FN & 546,190,291 \\
      PPV & 0.9810 \\
      Recall & 0.8488 \\
      TNR & 0.7334 \\
    \end{tabular}
    }
    \caption{Comparison of durations}
    \label{tab:accuracy_after_correction__merit}
  \end{subtable}
  \hfill %
  \begin{subtable}{0.45\textwidth} %
    \centering %
    \resizebox{0.65\columnwidth}{!}{
    \begin{tabular}{ r|r }
      & events \\
      \hline
      \rowcolor[HTML]{a0c080}
      True availability = TP &  72,787 \\
      \rowcolor[HTML]{C5E1A5}
      True outage = TN & 524 \\
      \rowcolor[HTML]{FFCCBC}
      False availability = FP & 210 \\
      \rowcolor[HTML]{FAD9D9}
      False outage = FN & 16,656 \\
      PPV & 0.9971 \\
      Recall & 0.8137 \\
      TNR & 0.7138 \\
    \end{tabular}
    }
    \caption{Comparison of events}
    \label{tab:accuracy_after_correction_events_merit}
  \end{subtable}
  \caption{Long-duration outages after precision-aware comparison for Durbin-with\merit (Dataset: 2021q1)}
  \label{tab:combined_accuracy_after_correction_merit}
\end{minipage}
\end{table*}

We compared seven days of data starting from 2021-01-10.
\autoref{tab:accuracy_after_correction__merit} shows the confusion matrix
 from this analysis
 when we define Trinocular outages as ground truth.

PPV is very good (0.9810),
  showing that Durbin-with-\merit's reported availability
  is almost always correct.
True negative rate (TNR) estimates how many outages are correct.
TNR (0.7334) is lower than PPV because outages are rare,
  so small differences between Durbin-with-\meritgen and
  Trinocular are noticeable.
Finally,
  recall (0.8488) is also quite good,
 indicating that Durbin successfully detects a high proportion of the outages identified by Trinocular. 

\subsubsection{Durbin-with-\merit, by events}
We next compare the number of outage events,
  comparing Durbin-with-\merit against Trinocular.
  
Similar to Durbin-with-\broot ,
  many blocks are always up, giving us good recall (0.8137).
A small fraction of blocks (about 10\%)
  are detected poorly in Durbin and produce many false events.
These blocks have sparse traffic,
  and when multiple consecutive timebins have no traffic,
  false outages result.
These sparse-traffic blocks reduce TNR.
As future work we plan to examine making timebin duration more adaptive
  to provide more reliable results for these blocks.
This problem has observed~\cite{richter2018advancing} before in active detection,
  where it was resolved by observing more data to confirm or reject the outage,
  a rough equivalent to increasing the timebin duration~\cite{Baltra20a}.

\subsection{Sensitivity of Belief Threshold}
\label{sec:sensitivity_beliefthreshold}

We next examine the sensitivity of our results to the belief threshold ($\theta_b$)
  an important parameter discussed in \autoref{sec:durbin_address_level}.
In our model, true outage detection varies with the change in belief threshold.
A higher belief threshold can guarantee not to get false reports on short gaps.
It also guarantees the detection of true outages.
But too high a threshold will miss very brief outages.
We customize parameters to find a middle ground to balance these two competing requirements.

\begin{figure*}[htbp]
\centering
\begin{minipage}{0.65\textwidth}
\begin{subtable}{0.5\textwidth}    
\centering
\resizebox{1.08\textwidth}{!}{ %
\begin{tabular}{ r|r|r|r } 
  & \multicolumn{3}{c}{\textbf{Belief Threshold ($\theta_b$)}}\\
  & 0.3 & 0.6 & 0.8 \\
 \hline
 \rowcolor[HTML]{a0c080}
 TA & 50,834,003,114 & 52,525,765,695 & 55,374,729,345 \\ 
 \rowcolor[HTML]{C5E1A5}
 TO & 16,964,568 & 13,147,965 & 12,934,568 \\ 
 \rowcolor[HTML]{FFCCBC}
 FA & 5,452,416 & 2,471,178 & 2,923,456 \\
 \rowcolor[HTML]{FAD9D9}
 FO & 2,347,645,465 & 78,163,261 & 87,853,262 \\ 
 PPV & 0.9999 & 0.9999 & 0.9999 \\
 Recall & 0.9558 & 0.9985 & 0.9984 \\
 TNR & 0.7567 & 0.8417 & 0.8191 \\
\end{tabular}
}
\caption{Durbin-with-\broot}
\label{tab:belief_threshold}
\end{subtable}
\hspace{1em} %
\begin{subtable}{0.48\textwidth}
\centering
\resizebox{1.05\textwidth}{!}{ %
\begin{tabular}{ r|r|r|r } 
  & \multicolumn{3}{c}{\textbf{Belief Threshold ($\theta_b$)}}\\
  & 0.3 & 0.6 & 0.8 \\
 \hline
 \rowcolor[HTML]{a0c080}
 TA & 3,264,854,345 & 4,908,712,735 & 4,345,576,243 \\ 
 \rowcolor[HTML]{C5E1A5}
 TO & 7,824,539 & 4,711,215 & 3,528,342 \\ 
 \rowcolor[HTML]{FFCCBC}
 FA & 5,026,362 & 1,826,015 & 1,124,462 \\
 \rowcolor[HTML]{FAD9D9}
 FO & 475,252 & 361,615 & 198,573 \\ 
 PPV & 0.9910 & 0.9914 & 0.9997 \\
 Recall & 0.9988 & 0.9981 & 0.9996 \\
 TNR & 0.6088 & 0.7206 & 0.7586 \\
\end{tabular}
}
\caption{Durbin-with-\merit.}
\label{tab:belief_threshold_merit}
\end{subtable}
\caption{Confusion matrix as belief threshold varies for Durbin-with-\broot and Durbin-with-\merit. Times in seconds.}
\label{tab:belief_threshold_merit_broot}
\end{minipage}
\hspace{1em} %
\begin{minipage}{0.32\textwidth}
\centering
\includegraphics[width=\textwidth]{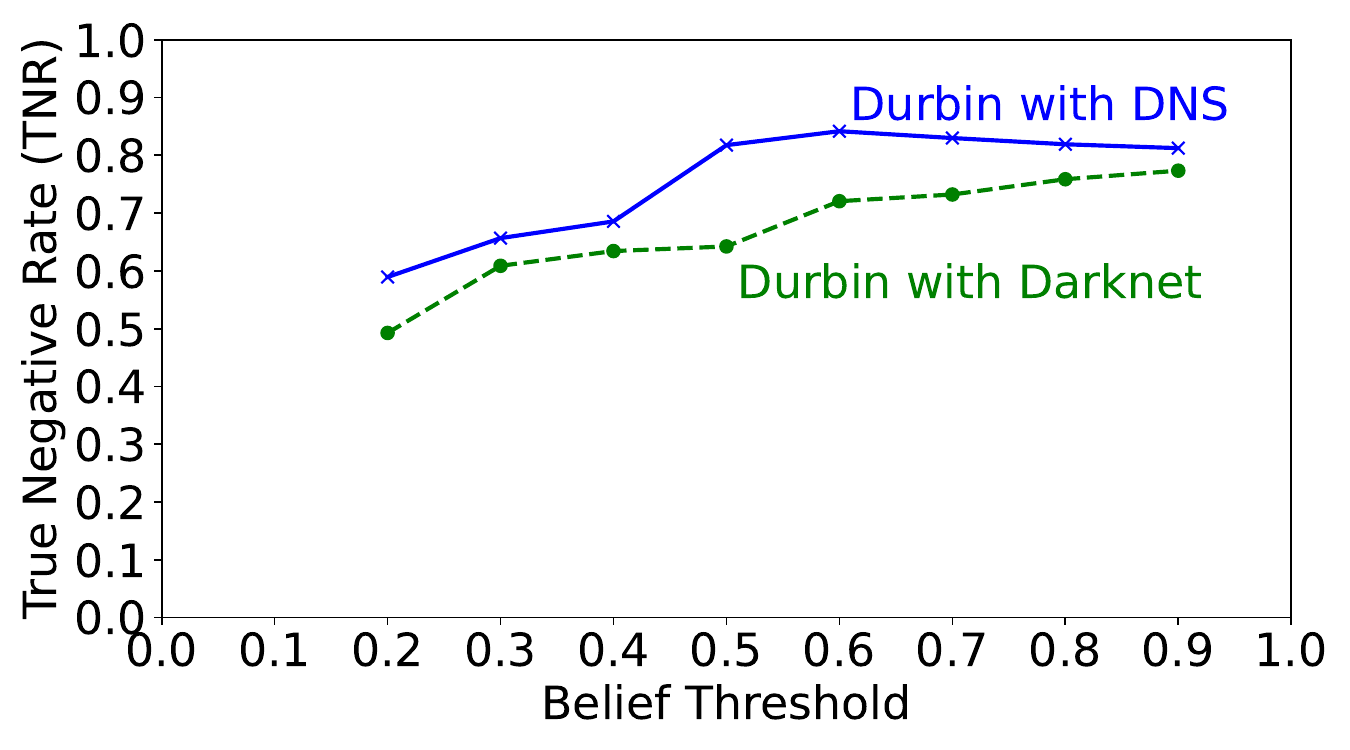}
\caption{True outage detection rate for Durbin-with-\broot and Durbin-with-\merit.}
\label{fig:beliefthreshold_vs_tnr}
\end{minipage}
\end{figure*}

To study the belief threshold we vary belief threshold $\theta_b$ and hold the time bin at $T(b) = 10$\,minutes.
We compare the impact of belief threshold $\theta_b$ of Durbin-with-\broot and Durbin-with-\merit against Trinocular,
  quantitatively (\autoref{tab:belief_threshold} and \autoref{tab:belief_threshold_merit})
  and graphically (\autoref{fig:beliefthreshold_vs_tnr}).

\autoref{tab:belief_threshold} and \autoref{tab:belief_threshold_merit} show the trend change by looking at parameters PPV, recall, and TNR with varying belief thresholds for 0.3, 0.6, and 0.8,
  and
  \autoref{fig:beliefthreshold_vs_tnr}
  compares TNR as threshold varies for more values between 0.2 and 0.9.

TNR is similar for threshold 0.6 or more.
We chose $\theta_b = 0.6$ to maximize sensitivity to short-duration outages.
We see the benefits of more sensitive detection 
  in \autoref{tab:belief_threshold}
  where $\theta_b = 0.6$ reports
  shorter outage duration (78.16\,Ms less of the 87.85\,Ms before)
  and larger true availability (52.52\,Gs duration than  50.83\,Gs before.)
  compared to  $\theta_b = 0.8$.

This evaluation also shows
  the cost of a low threshold.
False outages and false availability are higher
  with $\theta_b = 0.3$
  compared to thresholds of 0.6 or 0.8.
(For example,
  in \autoref{tab:belief_threshold},
  we see $2\times$ more false availability
  and $26\times$ more false outages.)
Both recall and TNR are lower for  $\theta_b = 0.3$
  because Durbin is detecting short inactive periods as outages.

\subsection{Sensitivity to Timebin Duration}
\label{sec:sensitivity_timebin}

\reviewfix{} Durbin optimizes parameters to provide temporal precision when possible,
  but falls back on coarser temporal precision when necessary to improve coverage and accuracy for both sparse and dense blocks.
\autoref{sec:sensitivity_beliefthreshold} compares PPV, recall and TNR
  as belief thresholds vary while setting $T(b)$ as constant. 
In this section, we will vary $T(b)$ and set the belief threshold as constant ($0.6$) to see the trend change in PPV, recall and TNR.

\autoref{tab:time_bin} shows Durbin's recall and TNR are very sensitive to timebin duration ($T(b)$).
Longer durations of $T(b)$ are more likely to miss short outages but improve coverage by including addresses with sparse data.

\begin{table*}[htbp]
\centering
\begin{minipage}{0.72\textwidth}
\begin{subtable}{0.52\textwidth} 
\centering
\resizebox{\textwidth}{!}{ %
\begin{tabular}{ r|r|r|r } 
  & \multicolumn{3}{c}{\textbf{Time bin duration ($T(b)$)}}\\
  & 15\,minute & 10\,minute & 5\,minute \\
 \hline
 \rowcolor[HTML]{a0c080}
 TA & 56,492,461,162 & 52,525,765,695 & 47,456,373,912\\ 
 \rowcolor[HTML]{C5E1A5}
 TO & 11,234,345 & 13,147,965 & 14,092,345\\ 
 \rowcolor[HTML]{FFCCBC}
 FA & 3,043,362 & 2,471,178 & 2,001,769\\
 \rowcolor[HTML]{FAD9D9}
 FO & 72,036,450 & 78,163,261 & 128,124,934\\ 
 PPV & 0.9999 & 0.9999 & 0.9999 \\
 Recall & 0.9999 & 0.9985 & 0.9973\\
 TNR & 0.7821 & 0.8417 & 0.8756 \\
\end{tabular}
}
\caption{Durbin-with-\broot}
\label{tab:time_bin}
\end{subtable}
\hspace{0.5em} %
\begin{subtable}{0.49\textwidth} 
\centering
\resizebox{\textwidth}{!}{ %
\begin{tabular}{ r|r|r|r } 
  & \multicolumn{3}{c}{\textbf{Time bin duration ($T(b)$)}}\\
  & 30\,minute & 20\,minute & 10\,minute \\
 \hline
 \rowcolor[HTML]{a0c080}
 TA & 4,974,983,931 & 4,908,712,735 & 3,064,574,810\\ 
 \rowcolor[HTML]{C5E1A5}
 TO & 4,025,307 & 4,711,215 & 9,833,309\\ 
 \rowcolor[HTML]{FFCCBC}
 FA & 2,021,539 & 1,826,015 & 1,615,440\\
 \rowcolor[HTML]{FAD9D9}
 FO & 267,706 & 361,615 & 546,190\\ 
 PPV & 0.9997 & 0.9914 & 0.9910 \\
 Recall & 0.9996 & 0.9981 & 0.9988\\
 TNR & 0.6656 & 0.7206 & 0.8588 \\
\end{tabular}
}
\caption{Durbin-with-\merit}
\label{tab:time_bin_merit}
\end{subtable}
\caption{Confusion matrix as $T(b)$ varies for Durbin-with-\broot and Durbin-with-\merit. Times in seconds.}
\label{tab:time_bin_merit_broot}
\end{minipage}
\hspace{1em} %
\begin{minipage}{0.23\textwidth}
\centering
\resizebox{\textwidth}{!}{ %
\begin{tabular}{ r|r } 
    & events \\
 \hline
 \rowcolor[HTML]{a0c080}
 True availability = TP & 31,115  \\
 \rowcolor[HTML]{C5E1A5}
 True outage = TN & 2,030  \\
 \rowcolor[HTML]{FFCCBC}
 False availability = FP  & 735 \\
 \rowcolor[HTML]{FAD9D9}
 False outage = FN & 1,799 \\
 PPV & 0.9769\\
 Recall & 0.9453\\
 TNR & 0.7341\\
\end{tabular}
}
\caption{Short-duration outages for Durbin-with-\broot (events)}
\label{tab:short_outages}
\end{minipage}
\end{table*}

\begin{figure*}[htbp]
  \centering
  \begin{subfigure}{0.32\textwidth}
    \centering
    \includegraphics[width=\textwidth]{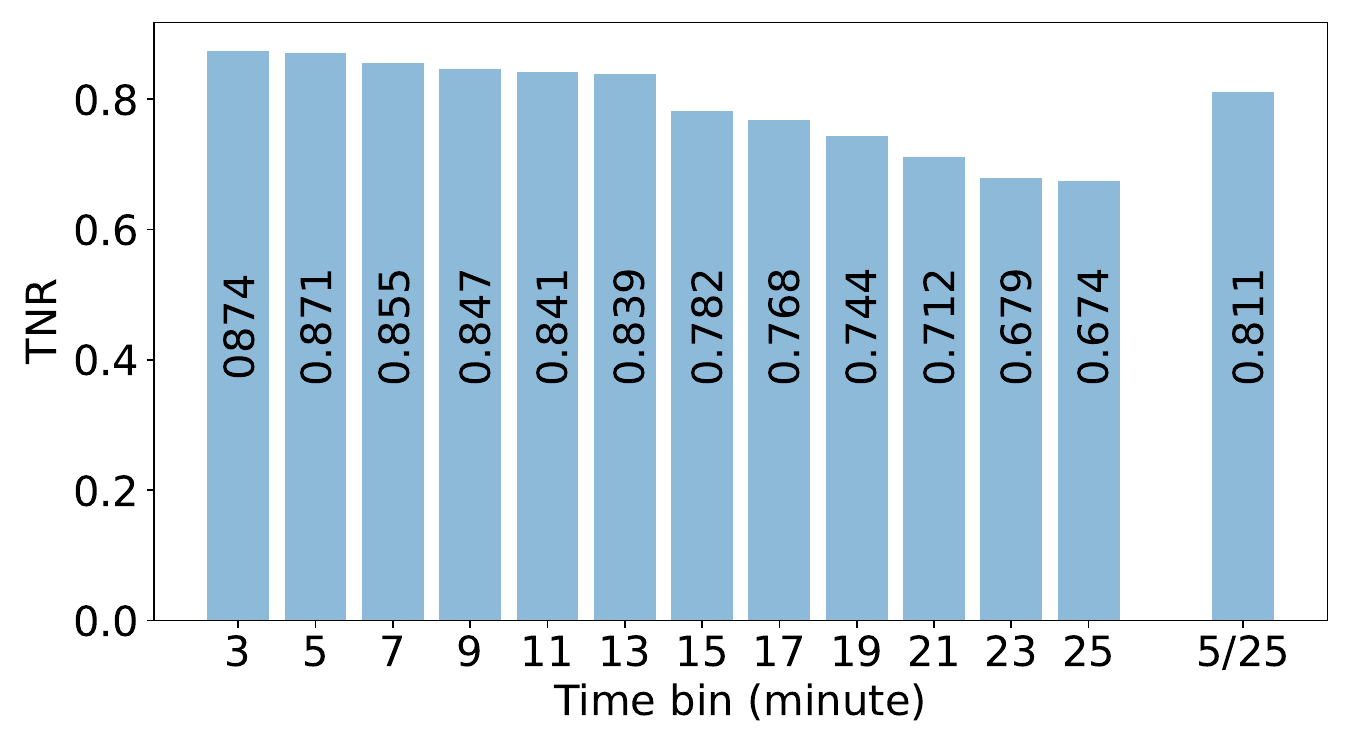}
    \caption{$T(b)$ vs.~TNR for Durbin-with-\broot}
    \label{fig:time_vs_TNR}
  \end{subfigure}
  \hfill
  \begin{subfigure}{0.32\textwidth}
    \centering
    \includegraphics[width=\textwidth]{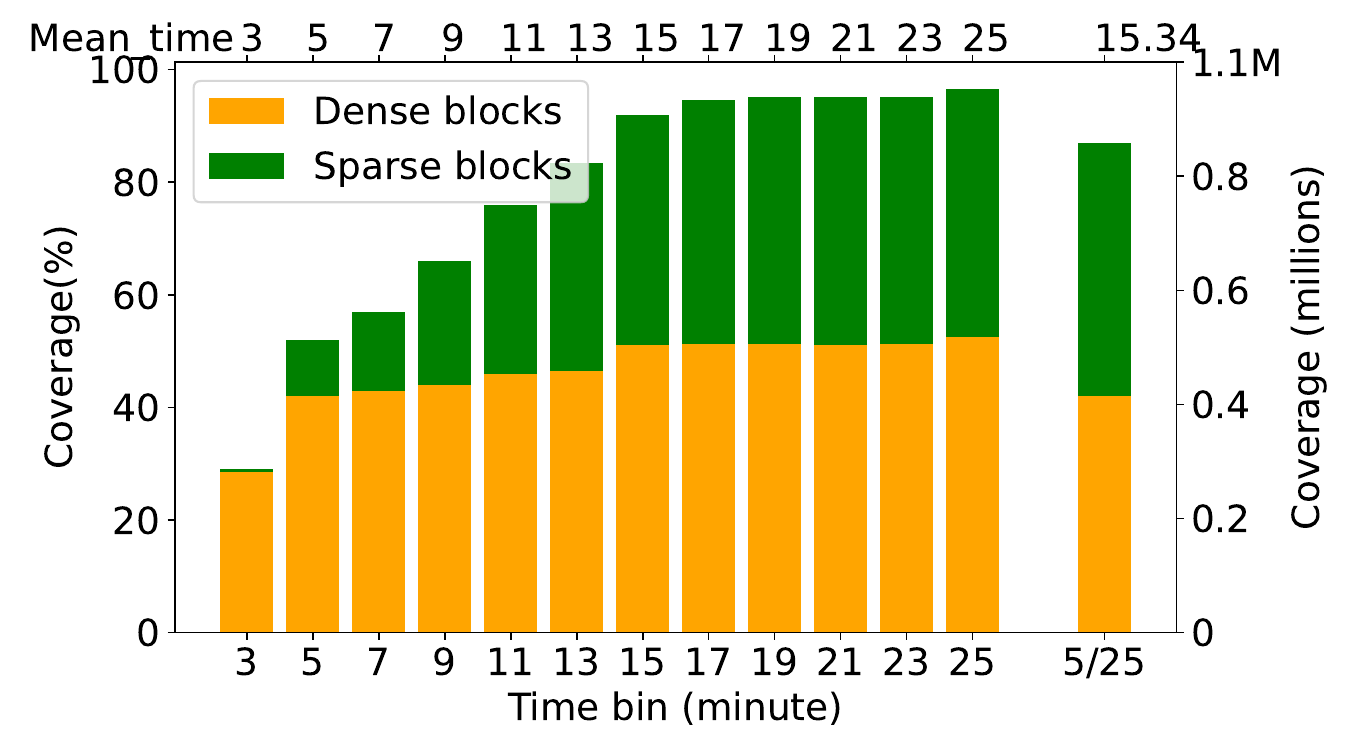}
    \caption{Tradeoff between $T(b)$ and coverage for Durbin-with-\broot}
    \label{fig:time_vs_coverage}
  \end{subfigure}
  \hfill
  \begin{subfigure}{0.32\textwidth}
    \centering
    \includegraphics[width=\textwidth]{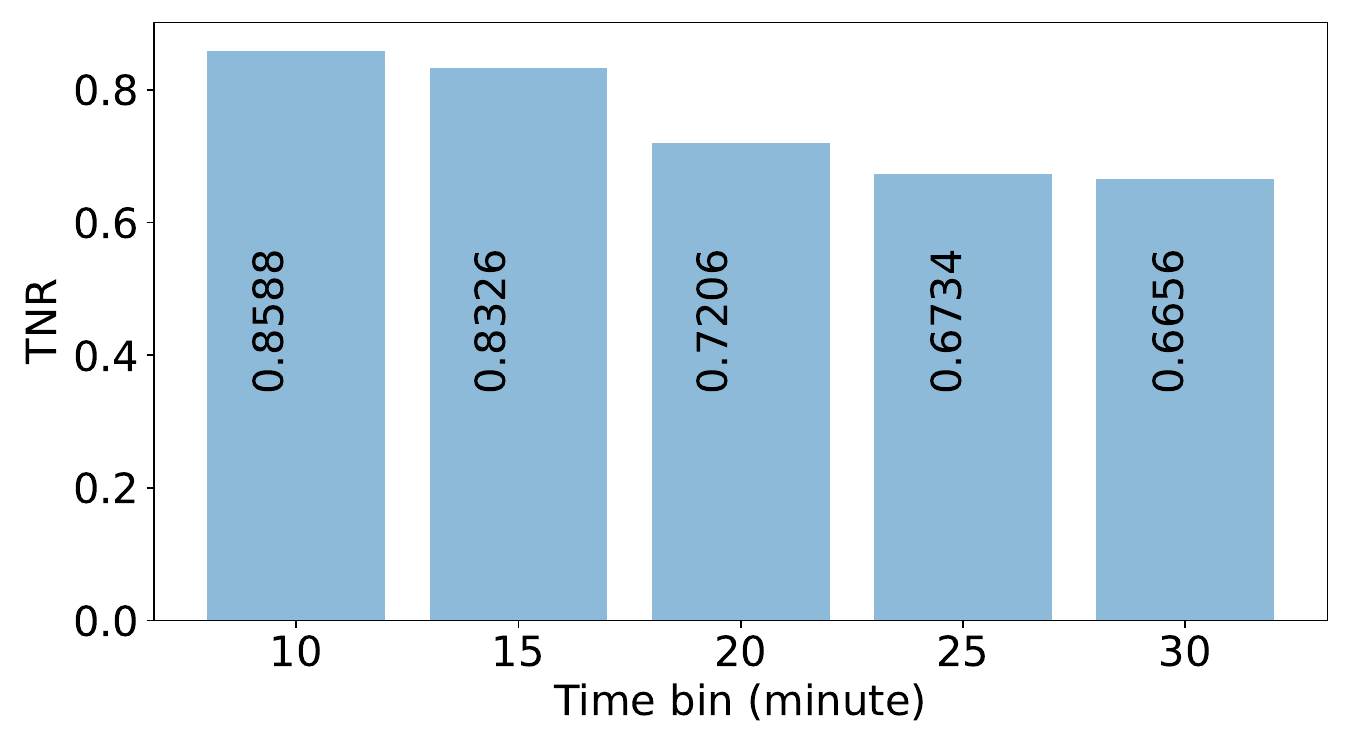}
    \caption{$T(b)$ vs.~TNR for Durbin-with-\merit}
    \label{fig:time_vs_TNR_merit}
  \end{subfigure}
  \caption{Sensitivity of the timebin for Durbin-with-\broot and Durbin-with-\merit.}
  \label{fig:trade_off}
\end{figure*}

On the contrary, short $T(b)$ can reduce coverage because it means that we are analyzing a smaller amount of traffic data at once making it more difficult to identify patterns or anomalies in the data, especially for sparse blocks or regions with low traffic volume. 
In \autoref{tab:time_bin} we show accuracy for three timebin durations in detail and in \autoref{fig:trade_off} we add more reference points and see the trend change of TNR and coverage.

\subsubsection{How does accuracy vary with timebin?}

Timebin duration is an important parameter to Durbin.
We next vary timebin duration's influence on accuracy
  to justify Durbin's choice of a $10$ minute timebin.
We hold the belief threshold constant at $\theta_b = 0.6$.

In \autoref{tab:time_bin},
  with a 10\,minute timebin for both sparse and dense blocks, the performance of PPV, Recall, and TNR is outstanding:
  0.9999, 0.9985, and 0.8417, respectively. 
In comparison, using a 5\,minute timebin
  reduces Recall slightly (0.9973).

Recall is lower for 5\,minute timebins
  because it increases the number of false outages.
We see this change as the false outage duration increases
  to 128.12\,Ms from 78.16\,Ms as we go to 5\,minutes from 10\,minute timebins.
\reviewfix{}
These false outages occur in blocks with infrequent traffic
  (\autoref{sec:accuracy}).
Although a shorter timebin results in some false outages,
  it also allows Durbin to detect previously missed short outage.
Durbin identifies one empty timebin without traffic as an outage when the timebin duration is short, even if there is no actual outage shown in \autoref{sec:case_studies}.

When timebin duration is longer,
  the true outage duration (in seconds) falls somewhat.
Again, this reduction is because some short-duration outages are missed.
\autoref{tab:time_bin} shows short timebins have more true outages (compare $T(b)$ of 15\,minutes vs 10 or 5).
Outage duration is reduced from 14.09\,Ms (when $T(b)$ is 5\,minute) to 11.23\,Ms (when $T(b)$ is 15\,minute).

Here we examined only outage duration.
In future work, we plan to evaluate counts of outage events in addition to durations.
\PostSubmission{}

We conclude that $T(b) = 10$ minutes provides the best choice
  when possible.
However, we select a longer duration from block with sparse traffic,
  as described in \autoref{sec:validation_trade-off}.

\subsubsection{How coverage varies with timebin?}
    \label{sec:coverage_vs_timebin}

The prior section examined three different timebins.
We next consider a wider range of options
  and study how TNR and coverage vary with timebin.

We can get both good coverage and TNR by setting short-duration timebin for dense blocks and long-duration timebin for sparse blocks.
We next experiment with different timebin durations
  to see how both sparse and dense blocks can achieve good coverage and TNR.

In \autoref{fig:time_vs_coverage} we vary $T(b)$ and study coverage.
(We hold the belief threshold constant at $\theta_b = 0.6$).
We vary $T(b)$ from 3\,minutes to 30\,minutes and observe the coverage.
Here we define coverage as the percentage of observed \broot blocks in a time bin with respect to the total number of existing \broot blocks.

Our coverage varies as a function of timebin because we pick coverage based on history discussed in \autoref{sec:durbin_history}.
In short-duration $T(b)$, we only observe dense blocks, which results in a high true outage detection rate (0.874) but low coverage (around 30\% dense blocks and 2\% sparse blocks of all measurable blocks). 
The short duration captures only a small amount of traffic in one $T(b)$, which excludes sparse blocks and leads to low coverage.

With long-duration $T(b)$ the coverage improves by including more sparse blocks, but the TNR suffers.

\autoref{fig:time_vs_TNR} shows TNR for timebin sizes from 3 to 25 minutes (left bars).
As timebin increases, more blocks become measurable and we increase coverage
  as we describe next.
When $T(b)$ is 25\,minute the coverage is around 95\% but the TNR is 0.674 because of the inclusion of sparse blocks.
By this analysis, we can say that Durbin can trade-off between spatial and temporal precision which is described in the next section \autoref{sec:validation_trade-off} 

\PostSubmission{}
In future work, we will perform two separate analyses: holding the coverage constant and observing changes in TNR as we varied $T(b)$ and allowing the coverage to vary while we monitored changes in TNR as we varied $T(b)$.
By comparing the effects of $T(b)$ under these two conditions, we can determine how much each factor contributes to changes in TNR and understand how timebin and coverage interact with each other.

\subsubsection{How accuracy varies with timebin, in Durbin-with-\meritgen?}

In this section, we show how varying the timebin duration influences outage detection performance, specifically in terms of metrics such as PPV, Recall, and TNR.
Shorter timebins improve the detection of shorter outages, especially for dense blocks, while longer timebins enhance reliability for sparse blocks by reducing false positives.
In \autoref{tab:time_bin_merit},
  with a 10\,minute timebin for both sparse and dense blocks, the performance of PPV, Recall, and TNR is outstanding, achieving values of 0.9910, 0.9988, and 0.8588, respectively. 

When timebin duration is longer, we observe a reduction in the true outage duration (in seconds), indicating that some outages are missed.
Similar to \autoref{tab:time_bin}, \autoref{tab:time_bin_merit} shows short timebins have more true outages (compare $T(b)$ of 30\,minutes vs 20 or 1).

Therefore, if we have a fixed timebin then 20\,minutes time bin seems good for Darknet.
We can see around $5\%$ more true outages when the time bin is 20\,minutes.
But if we have a variable timebin then different timebins for dense and sparse blocks are good \autoref{sec:validation_trade-off}.

\subsection{Trading Between Spatial and Temporal Precision}
\label{sec:validation_trade-off}
We exploit the ability to trade-off between spatial and temporal precision
  while preserving accuracy.
We customize parameters to treat each block differently, allowing different regions to have different temporal and spatial precision.
As a result, we can get coverage in sparse blocks,
  although to get good accuracy we must use coarser temporal precision.

In \autoref{fig:time_vs_coverage}
 we evaluate this trade-off,
 showing that we have fine precision for the dense blocks,
 but require coarser precision to cover blocks with sparse traffic
 across the left bars.
The rightmost bar, labeled 5/25,
  shows a hybrid system
  where blocks with dense traffic use $T(b)$ of 5\,minutes,
  while those with sparse traffic have $T(b)$ of 25\,minutes.

With varying $T(b)$ values (customized to block traffic),
  we obtain broad coverage: 85\% of all blocks with \broot traffic.
Varying $T(b)$ by block also provides a good TNR (0.811).
By contrast, a strict 25\,minute $T(b)$ has TNR 0.647,
  because coarser precision can miss short outages.
Comparing the rightmost two bars in \autoref{fig:time_vs_TNR},
  this is about a 20\% improvement in TNR.

This comparison shows the advantage of tuning parameters to each block
  to maximize coverage \emph{and} accuracy.

\subsection{Can We Detect Short-Duration Outages?}
\label{sec:validation_shortoutages}

We next demonstrate that Durbin can detect shorter outages than prior systems, in part because we can trade-off spatial and temporal precision with coverage and accuracy.
Here, we examine 5\,minute outages, with a belief threshold of 0.6 and $T(b)$ to 5\,minute.

To validate our short-duration outage results, we compare them to Disco~\cite{shah2017disco} using RIPE Atlas data~\cite{staff2015ripe} as ground truth.
Using RIPE data, Disco infers that multiple concurrent disconnections
  of long-running TCP connections in the same AS indicate a network outage.

Although its coverage is only about 10k /24s,
  we use Disco for ground truth to compare short outages
  because it reports 5\,minute outages
  (unlike Trinocular's 11 minutes).

  \PostSubmission{}
We study all 10.5k /24 blocks observed from both Durbin using \broot data
  that also have data from RIPE Atlas
  over 7 days of data starting on 2019-01-09.
\autoref{tab:short_outages} shows the confusion matrix,
  testing Durbin against Atlas disconnections as ground truth.
  Our model can correctly detect outages with short lengths which can be as little as 5 minutes or less than 5\,minutes.
When routing changes there can be transients on the internet which can cause brief outages.
We observe that we have great PPV (0.9769), recall (0.9453) and TNR (0.7341) for short-duration outages (5\,minutes or more). 
Our measurements show that on that week,
  around 5\% of total blocks that have 5\,minute outages that were
  not seen in prior work.
The duration of outages from 5 to 11\,minutes, omitted from previous observations, increases total outage duration by 20\%.

\subsection{Are These Results Stable?}
\label{sec:validation_long-term}

We next validate the consistency of our results over a week, examining accuracy with PPV, recall, and TNR for long-term observation to show the results are stable. 
We run Durbin on \broot data everyday for continuous monitoring. 

\subsubsection{Observation for a week}
In \autoref{fig:accuracy_7days} we observe seven days of Durbin's accuracy parameters (PPV, recall and TNR) (We put the figure in appendix \autoref{sec:appen_stable}).
PPV is the same for all seven days (0.9999) because of the high true availability on each day.
Recall and TNR are also generally stable,
  with recall ranging from 0.96 to 0.99 and TNR from 0.8 to 0.9.
Both are lowest on 2019-01-11, because on that day we see many sparse blocks which gives more false outages.

This data suggests that our results are consistent over multiple days.

\subsubsection{Continuous Monitoring}

As Durbin matures, we are shifting to running it continuously.
Thus far we have revised our implementation to
  run against \broot continuously.
We currently run it once at the end of the day.
It generates and caches training data for the current day,
  and then runs detections on the current day using
  cached training data from the prior two days.
Thus far
  we have run Durbin-over-\broot
  against a full month of data,
  and we are currently bringing up 24x7 processing.

Since our IPv6 \broot data mirrors the structure of IPv4 \brootgenshort,
  the validation of Durbin performance for IPv4
  gives confidence that IPv6 accuracy will be similar.

\section{Results}
\label{sec:results}

Having established that Durbin works in \autoref{sec:validation},
  we next explore what it says about the Internet.
\reviewfix{}
Our results show the outage rate on IPv4 and IPv6 in \autoref{sec:IPv6outage} and our IPv6 coverage in \autoref{sec:IPv6coverage}.

\subsection{How Many IPv6 Outages?}
\label{sec:IPv6outage}

We evaluate IPv6 outage rate based on seven days of
  passive data from \broot.
We established Durbin's accuracy in IPv4 (\autoref{sec:accuracy}), and showed that IPv6 sources
  have a similar traffic rate as IPv6  (\autoref{sec:traffic_per_address}).
  \PostSubmission{sigh, I don't see that 6.1 (how many ipv6 outages) compares against PRIOR WORK.  But that would be a good addition to that section, as a new subsection.
    will add this in future. -asma}
Using this result, we next provide the first results for IPv6 outages.

\begin{figure*}[htbp]
  \centering
  \begin{minipage}{0.43\textwidth}
    \centering
    \includegraphics[width=0.7\textwidth]{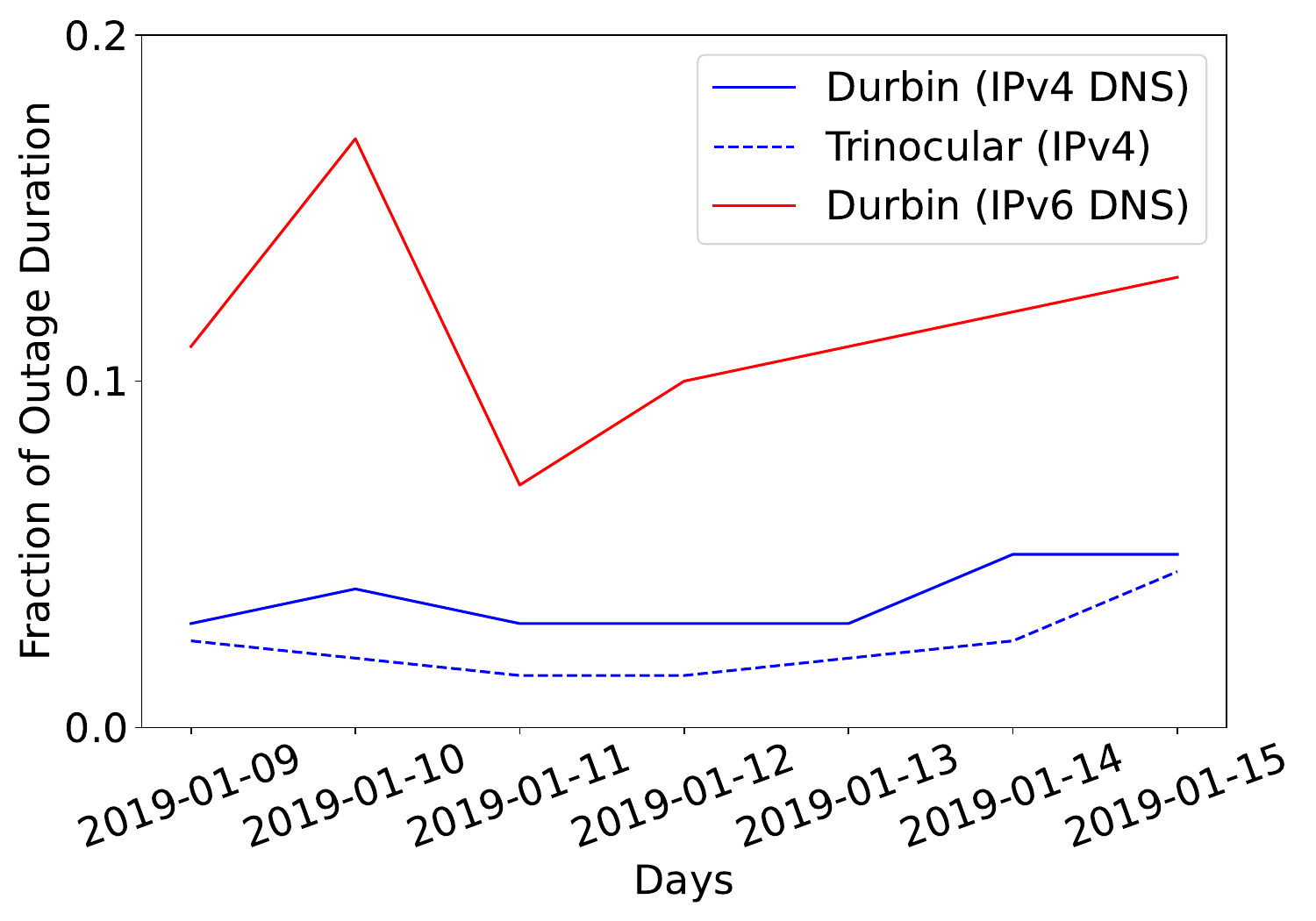}
    \caption{IPv4 and IPv6: outage fraction}
    \label{fig:blocks_outagefraction}
  \end{minipage}
  \hspace{1em}
  \begin{minipage}{0.53\textwidth}
    \centering
    \includegraphics[width=0.6\textwidth]{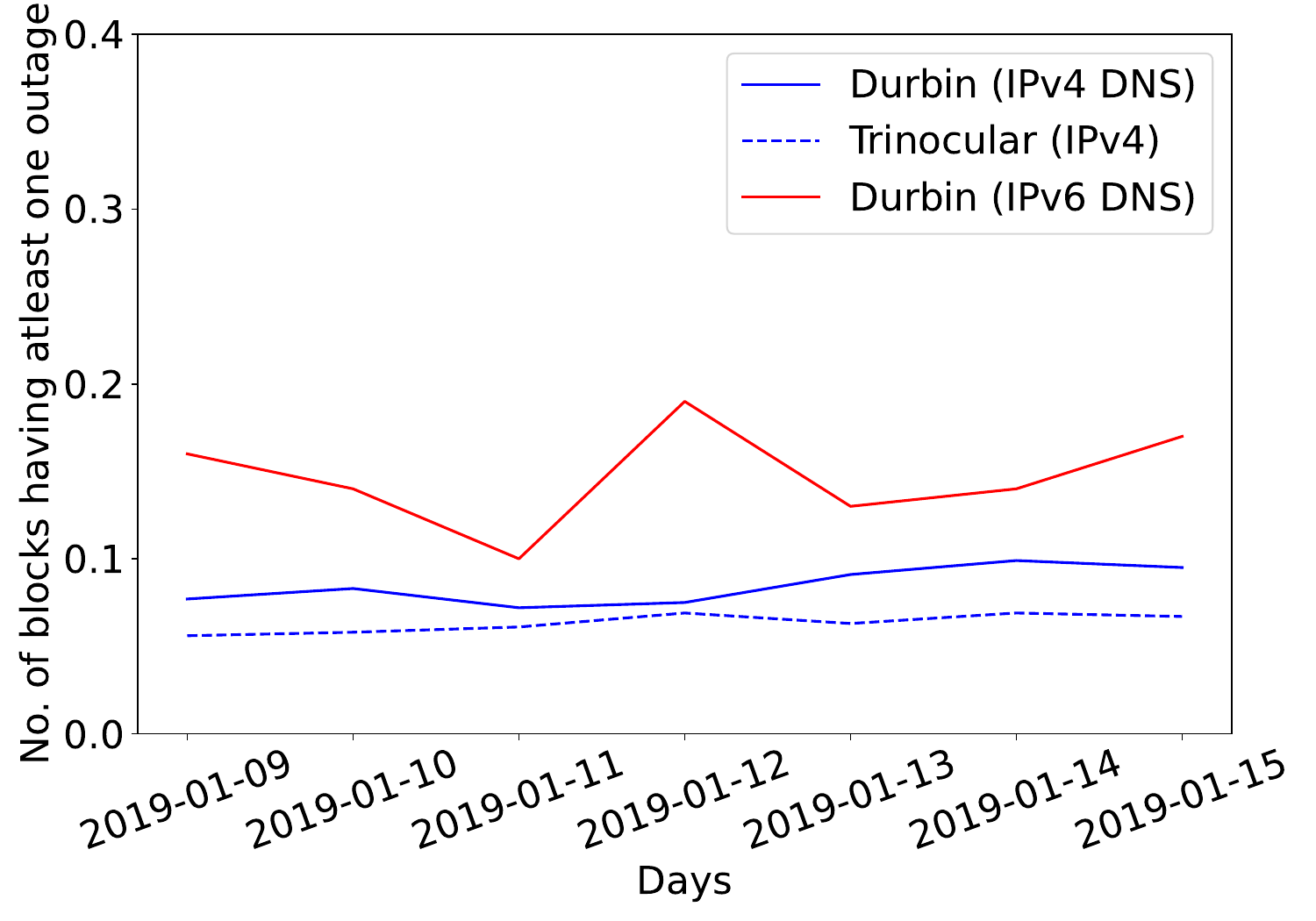}
    \caption{IPv4 and IPv6: blocks with at least one outage}
    \label{fig:blocks_onoutage}
  \end{minipage}

\end{figure*}

\begin{figure}
    \centering
    \includegraphics[width=0.9\columnwidth, trim=50 20 20 30]{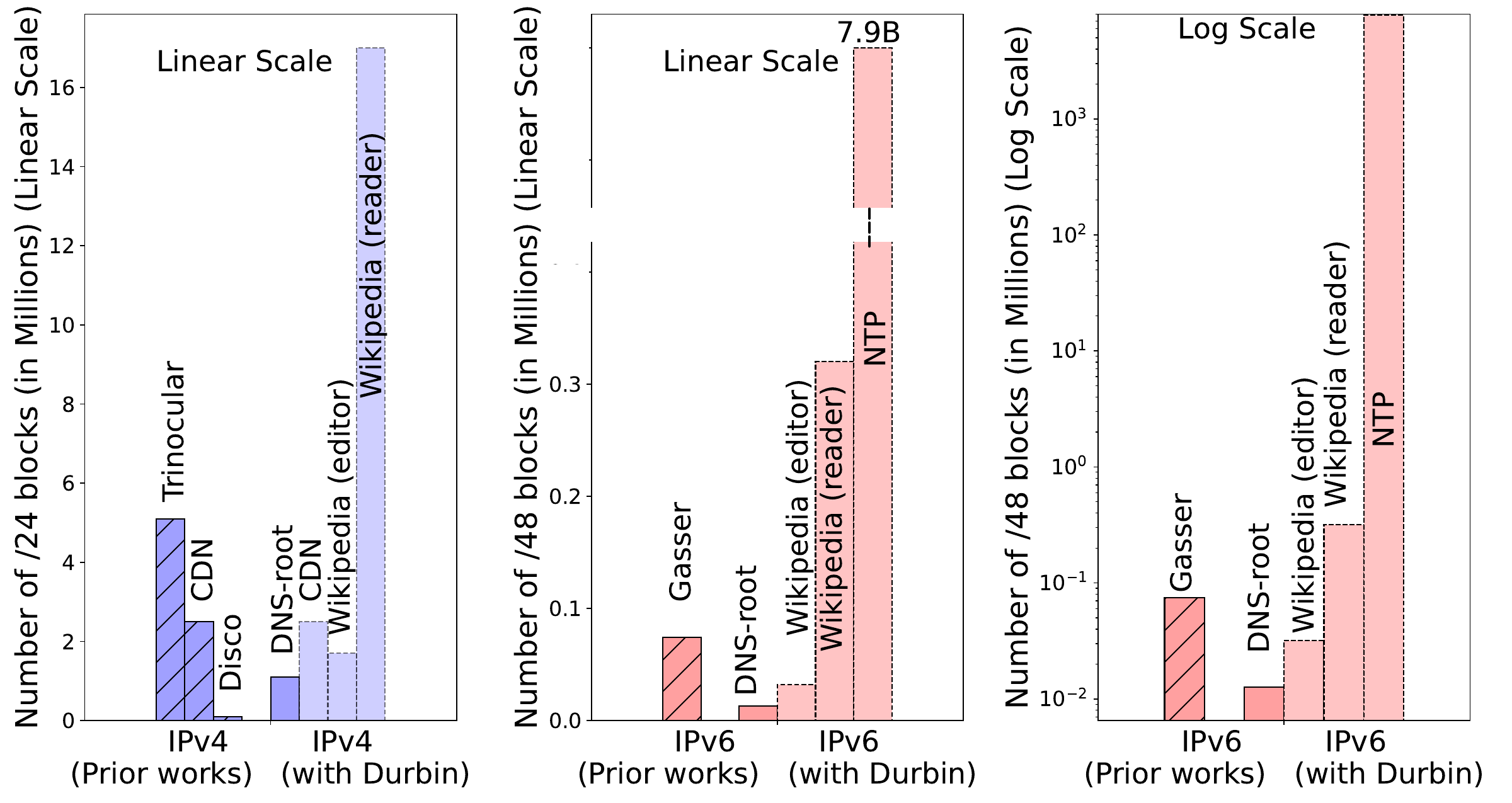}
    \caption{Current (dark shades) and potential (light) coverage in IPv4 and IPv6 for prior work (hashed) and Durbin (solid).}
    \label{fig:ipv6_coverage}
\end{figure}

\subsubsection{Outage Duration}

First, we show internet outage duration (seconds of outages for fraction of time for outages) of IPv4 and IPv6 address blocks and Trinocular's outage \autoref{fig:blocks_outagefraction} for seven days.
The fraction of outage duration follows prior work
  and allows us to compare IPv4 outages against IPv6.

In \autoref{sec:validation_shortoutages} see /48 IPv6 blocks are out about 9\% of the time.

To show that both Durbin and Trinocular have similar outages in IPv4, we compare Durbin's outage duration with Trinocular's outage duration for IPv4 blocks. 
By comparison, Durbin sees around 1\% outage duration in /24 IPv4 blocks.
Durbin seems to have a bit higher outage duration than Trinocular,
  due to its detection of short outages (see \autoref{sec:validation_shortoutages}).

We first examine how often Durbin infers networks are down.

We evaluate a week of data using Durbin-with-\broot
  in \autoref{fig:blocks_outagefraction}.
Here the Durbin-inferred fraction of outages
  are shown for IPv4 (blue, solid line)
  and for IPv6 (the red line).
We compare to the publicly shared outage rate from Trinocular for IPv4
  (the blue dashed line).

First, we see that IPv4 outage fractions are similar
  for both Durbin and Trinocular, both around 2\% to 4\%.
Durbin's results are from 1M /24 IPv4 blocks,
  less coverage than Trinocular's 5M, but a similar fraction.

Durbin also provides
  \emph{the first data for IPv6 outages}
  (the red line).
We see IPv6 outages are much higher: from 0.6 to 1.5,
  rates $10\times$ what we see in IPv4.
This data is a report of 1338 outages in /48 IPv6 blocks.
The absolute number of outages is significantly higher for IPv4 (1M) compared to IPv6 (30k) due to the larger number of measurable IPv4 blocks.

We expect that IPv6 outages will be more than IPv4,
  but evaluating why the difference is this large is ongoing work.
  \PostSubmission{}
\reviewfix{}
Recent work that studied outage rates in RIPE Atlas~\cite{salujadifferences},
  finding that IPv6  DNS query rates fail at a rate $3\times$ what IPv4 sees
  (about 9\% failures in v6 vs. 3\% in v4).
They point to a combination of end-system misconfiguration
  and long-term peering disputes as reasons for the difference.
Our results show a larger difference, something we are looking into.

\subsubsection{Blocks With At Least One Outage}

We next consider how many blocks see at least one outage each day.

We consider this metric because
  outage durations are heavily influenced by a few blocks that are unreachable
  for a long time.
Blocks with long-term outages are unlikely to have active users,
  but users remember times their work was stopped by an Internet outage---something captured by this metric.

In \autoref{fig:blocks_onoutage}
  shows the fraction of blocks with at least one 10-minute outage on each day.
For IPv4,
  Durbin sees 3 to 4\% of /24 IPv4 blocks
  have at least one outage on any given day (the solid blue line).
Trinocular shows a similar fraction, with about 2\% of blocks out at least once (the dashed blue line),
  a similar fraction as what Durbin sees.
For IPv6, that fraction rises to 12\% to  13\% of all measurable /48 IPv6 blocks
  (the solid red line).

We can use this metric to compare IPv4 and IPv6:
  we see that the outages for IPv6 seems somewhat greater than for IPv4,
  suggesting IPv6 reliability can improve.
The absolute number of outages is much larger for IPv4 (1M) than for IPv6 (30k) because
  there are many more measurable IPv4 blocks.
  
%
%
%
%
%
%
%
%

%
%

%
%
%
%
%
%

\subsection{How Broad Is Our IPv6 Coverage?}
\label{sec:IPv6coverage}

We next evaluate Durbin's current coverage
  along with Durbin's \emph{potential} coverage
  given access to other data sources,
  comparing both to the prior work.

Durbin's coverage depends on its input data---any data source that
  supports the passive observation of global sources is suitable input (\autoref{sec:data_requirements}).
Potential coverage is maximized using data sources that see many source addresses.
Most top-10 websites (Google, Facebook, Wikipeida, etc.)
  more than meet this requirement,
  as would many global CDNs (Akamai, Amazon Cloudfront, Cloudflare, etc.),
  and some global services (public DNS resolvers, NTP services, etc.).
\reviewfix{}
While we do not currently have access to this data,
  and it is unlikely a commercial service would share their data with
    researchers,
  we consider potential Durbin coverage to show
    how well the method \emph{could} work, given proper input.
(It is always possible that a top website or CDN would choose to implement
  Durbin for their own purposes.)

Published work shows Wikipedia sees 25M unique IP addresses~\cite{tran2013cross},
  which suggests they likely see millions of /24 IPv4 prefixes.
Analysis of ``a major CDN'' states they have 2.3M ``trackable'' /24s address blocks
  (where trackable means blocks with at least 40 active IP addresses,
  allowing outage detection by their method).
NTP sees \emph{billions} of IPv6 addresses~\cite{rye2023ipv6}.  
Below we evaluate \broot, a CDN, Wikipedia, and NTP as potential data sources for Durbin.

\subsubsection{Comparing Durbin Coverage to Prior Work:}

%


%
\autoref{fig:ipv6_coverage}
  compares prior systems (left, hashed bars)
  against Durbin 
  with \broot (the first darker blue bar in the right cluster).
We show data for both IPv4 (the left graph) and IPv6 (the right graph),
  normalizing both to 100\% as the best possible current result.

For IPv4 (the left graph with blue bars),
  we see that Durbin with \broot provides good coverage:
  about 1M /24s blocks.
We find this coverage surprisingly good, given \broot only sees traffic
  from DNS servers, not end-users.
Durbin's IPv4 coverage   
  is about one-fifth of the 5.1M in Trinocular (the largest current outage detection system),
  half of CDN-based detection,
  and $10\times$ more than Disco.

Durbin's \emph{current} coverage is determined by \broot,
  and which /24 blocks report traffic that is frequent enough to evaluate
  (\autoref{sec:durbin_history}).
We evaluate Durbin's IPv6 coverage based on one representative day (2019-01-10) of
  passive data from \broot,
  comparing results in IPv4 and IPv6.

We evaluate Durbin's IPv6 coverage based on one representative day of
  passive data from \broot,
  comparing results in IPv4 and IPv6.
We show Durbin's coverage of both IPv4 and IPv6 address blocks and compare the coverage with prior works Trinocular, Akamai and Disco.
We use Trinocular, Akamai, and Disco coverage as the prior work for IPv4 and Gasser hitlist for IPv6 coverage in \autoref{fig:ipv6_coverage}.

\subsubsection{Durbin IPv4 Coverage with Potential Alternative Sources:}

Durbin's current coverage is limited by not seeing clients,
  but if Durbin were run with a major website's logs as input,
  its coverage can equal or exceed current systems.
We next consider what Durbin coverage \emph{would} be
  if it was applied to CDN, Wikipedia, or NTP traffic.

\textbf{CDN:}
We estimate potential coverage with CDN data from
  published work~\cite{richter2018advancing}.
However, their paper provides only IPv4 coverage.

\textbf{Wikipedia:}
We use Wikipedia as an example top-10 website.
Wikipedia does not provide public access to browsing traffic,
  but all Wikipedia edits are public, and about half are logged with IP addresses
  (as disclosed to the editor, so with their consent).

We downloaded the entire Wikipedia edit history and
  the logging history and extracted all ``IP users''.
We count 9,264,603 unique IPv4 addresses and  94,042 IPv6 addresses,
  for 1,694,599 IPv4 /24 blocks and  30,257 IPv6 /48 blocks.

Of course, Wikipedia has \emph{far} more readers than editors.
One analysis observes that although the read: edit ratio is not known,
  the number of page views can provide an upper bound on the number of readers.
All Wikipedia sees about 85 billion page views per month,
  and Hill suggested 35 page-views per reader~\cite{Hill11a}, implying about 686M readers per month,
  or 75$\times$ more than editors.
We suggest this offers a loose \emph{upper bound}
  on the number of unique IPs that Wikipedia sees per month.
We assume a more conservative $10\times$ multiplier from editors,
  implying readers will show 1.7M IPv4 /24 blocks and 30.2k IPv6 /48 blocks.

\textbf{Implications for Durbin:}
IPv4 coverage with the CDN will roughly match coverage with prior work~\cite{richter2018advancing},
  but Durbin will be able to report 5-minute temporal precision for frequent-traffic blocks.
This analysis uses only blocks reported as measurable by their outage detection system.
It is possible that Durbin could provide coarse-time results for blocks that are unmeasurable by their method, thereby increasing coverage.

These results suggest that the Durbin algorithm could provide
  at least as good coverage as CDN-based outage detection,
  when applied to a data source like a major website.

\subsubsection{Actual and Potential IPv6 Coverage:}
Durbin coverage is even more promising when one considers IPv6.
Here we compare the IPv6 hitlist as the best possible option
  (although we have not seen published work using IPv6 hitlists for outage detection).
We cover slightly less than one-fifth of the Gasser hitlist,
  but use only passive data from \broot.

Our analysis of Wikipedia IPv6 edits suggests Durbin would see 30.2k /48 blocks,
  roughly double \broot, and half of the Gasser IPv6 hitlist.
Projecting edits to readers with the same ratio as in IPv4, we expect around 300k /48 blocks.

Finally,
  while our analysis of Wikipedia is conservative,
  Rye and Leven took data from NTP,
  a global service touched by billions of IPv6 addresses.
The right-most set of red bars in \autoref{fig:ipv6_coverage}
  add NTP, but with a \emph{log-scale} $y$-axis:
  with 7.9 billion IPv6 addresses,
  NTP exceeds all other sources.

%
%
%
%
%




%
%
%
%
%

%
%


\section{Conclusion}

We have describe Durbin, a system
  to detect Internet outages
  with a new adaptive algorithm
  using passive sources.
The challenge to outage detection from passive
  data is balancing accuracy with spatial and temporal precision and coverage;
Durbin
  provides good accuracy (0.811 TNR)
  at constant spatial precision (/24 IPv4 and /48 IPv6 blocks)
  by adapting temporal precision for each block (5 or 25\,mintues).
We evaluated Durbin with two different data sources:
  \broot and \merit,
  examples representing \brootgenlong and \meritgens.
Coverage of IPv4 with our data sources is good (about 1M /24 blocks with \broot).
IPv4 coverage with \broot is large as current active methods,
  but a top website could use Durbin to see IPv4 coverage equal to or exceeding active methods.
Finally, Durbin provides the first published data reporting IPv6 outages (30k /48 IPv6 blocks with \broot),
  showing the promise passive methods to track outages in IPv6.

\label{page:last_body}

\bibliographystyle{plain} 

%
\begin{appendices}
\appendix

\section{Research Ethics}
	\label{sec:research_ethics}

Our work poses no ethical concerns.
In evaluating the risks of or work relative to its benefits,
  it poses minimal risks,
  while there are significant benefits to a new method to detect Internet outages
  and thereby improve Internet service.

We believe our work poses minimal risk because
  our approach analyzes passive traffic
  to look for activity on networks.
The primary risk is that such traffic analysis may reveal personal information
  about individuals.
To avoid revealing such information,
  our data provider provides only an IP address and timestamp of activity,
  not the actual user activity.
(For example, in DNS data, the query and reply are removed.)
We discuss data handling in \autoref{sec:data_requirements}:
  while we require tracking specific sources,
  we do not need to know actual IP addresses, just correct address blocks.
Our data provider therefore anonymizes the least-significant bits of IP addresses.

Finally, our use of \broot poses minimal privacy risk because
  nearly all DNS queries are from infrastructure (recursive resolvers),
  not directly from individuals,
  any requests from individuals are mixed in with queries from infrastructure.

We have submitted an IRB review request proposing an analysis
  of new data sources (beyond \broot) with the above anonymization
  as non-human-subjects research.
This IRB is currently under review.

\section{Are These Results Stable?}
	\label{sec:appen_stable}

\begin{figure}
\centering
\includegraphics[width=1.0\linewidth]{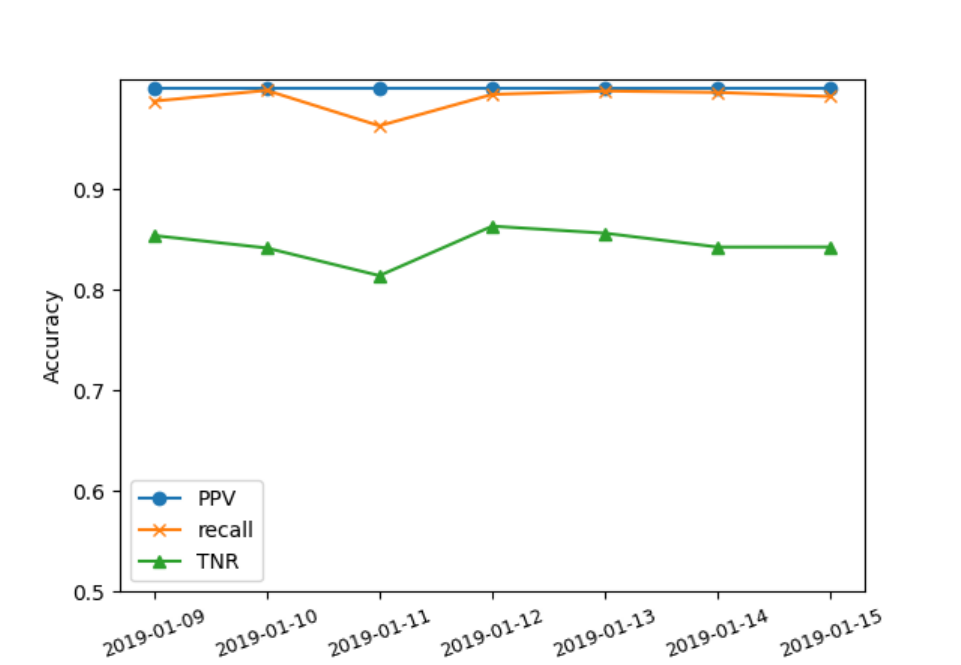}
\vspace{-0.25in}
\caption{PPV, recall and TNR for seven days for Durbin-with-\broot}
\label{fig:accuracy_7days}
\end{figure}

In \autoref{fig:accuracy_7days}, we observe seven days of Durbin's accuracy parameters (PPV, recall, and TNR). 
PPV is consistent for all seven days (0.9999) because of the high true availability on each day.
Recall and TNR are generally stable, with recall ranging from 0.96 to 0.99 and TNR from 0.8 to 0.9.
Both are lowest on 2019-01-11 due to many sparse blocks causing more false outages.

\end{appendices}
\label{page:last_page}

\end{document}